\documentclass[letterpaper,twocolumn,10pt]{article}
\usepackage{usenix}
\usepackage{color}
\usepackage{multirow}
\usepackage{xurl}
\usepackage{algorithmicx,algorithm}
\usepackage[noend]{algpseudocode}
\usepackage{subfig}
\usepackage{array}
\usepackage{colortbl}
\usepackage{enumitem} \usepackage{times}
\usepackage{xspace}
\usepackage{latexsym}
\usepackage{amsmath}
\usepackage{booktabs}
\usepackage{tabularx}
\usepackage{balance}
\usepackage{graphicx}
\usepackage{svg}
\usepackage{pifont}
\usepackage[font=small]{caption}
\svgsetup{
    inkscapepath=i/svg-inkscape/
}
\svgpath{{svg/}}
\usepackage{amssymb}
\usepackage{amsmath}
\usepackage{etoolbox}
\usepackage[all]{nowidow}

\setlist{nolistsep}
\newcommand{\parabf}[1]{\medskip\noindent\textbf{#1}}

\newcommand{\cut}[1]{}

\newcommand{\sysname}{HydraServe\xspace}

\begin{document}
\sloppy
\date{}

\title{\sysname: Minimizing Cold Start Latency for Serverless LLM Serving \\ in Public Clouds}

\pagestyle{plain}

\author{
\rm{Chiheng Lou$^{1}$ \enskip
    Sheng Qi$^{1}$ \enskip
    Chao Jin$^{1}$ \enskip
    Dapeng Nie$^{2}$ \enskip
    Haoran Yang$^{2}$ \enskip
    Yu Ding$^{2}$ \enskip 
}
\vspace{1mm}
\\
\rm{
    Xuanzhe Liu$^{1}$ \enskip
    Xin Jin$^{1}$ \enskip
}
\vspace{3mm}
\\
{
$^{1}$\textit{School of Computer Science, Peking University} \hspace{1.2em}
$^{2}$\textit{Alibaba Group}
}
}

\maketitle

\begin{abstract}
\label{sec:abstract}

With the proliferation of large language model (LLM) variants,
developers are turning to serverless computing for cost-efficient LLM deployment.
However, public cloud providers often struggle to provide performance guarantees for serverless LLM serving due to significant cold start latency caused by substantial model sizes and complex runtime dependencies.
To address this problem, we present \sysname,
a serverless LLM serving system designed to minimize cold start latency in public clouds.
\sysname proactively distributes models across servers to quickly fetch them,
and overlaps cold-start stages within workers to reduce startup latency.
Additionally, \sysname strategically places workers across GPUs to avoid network contention among cold-start instances.
To minimize resource consumption during cold starts,
\sysname further introduces pipeline consolidation that can merge groups of workers into individual serving endpoints.
Our comprehensive evaluations under diverse settings demonstrate that \sysname reduces the cold start latency by 1.7$\times$-- 4.7$\times$ and improves service level objective attainment by 1.43$\times$--1.74$\times$ compared to baselines.
\end{abstract}
 \section{Introduction}
\label{sec:introduction}

Large language models (LLMs) have become an essential part of daily work and life, powering a wide range of applications such as chatbots~\cite{gpt4, claude, claude2}, programming assistants~\cite{copilot, cursor, qdeveloper}, and AI-enhanced search engines~\cite{newbing, chatgpt-browsing}.
While the industry is releasing numerous LLMs with varying structures, sizes, or training corpora for diverse scenarios~\cite{opt, gpt3, gpt4, claude, claude2, falcon, gemma, gemma2, palm, llama, llama2, gemini, mistral, phi3, deepseek-v3},
businesses of different scales are also seeking to develop custom LLMs tailored to their unique needs~\cite{iclr22_lora, blog_botpnguin, blog_kili, blog_azure}.
As a result, a global ecosystem comprising over one million models has emerged, as documented by Hugging Face~\cite{huggingface}.

To efficiently deploy these models, serverless LLM serving has emerged as a compelling choice~\cite{huggingface, sagemaker, fc-gpu, kserve, nuclio, serverlessllm, blitzscale,lambdascale, deepflow}.
In serverless inference, the cloud platform automatically provisions GPUs for models based on the actual request load,
and model owners are only charged for the resource consumption during request processing.
This approach offers significant cost savings for long-tail models with infrequent usage.

However, for cloud providers, delivering performance guarantees for serverless LLM users is challenging, primarily due to the long latency of cold starts, i.e., provisioning new workers to meet the current load.
Production traces reveal that bursty traffic patterns are prevalent in both LLM serving workloads~\cite{burstgpt} and serverless workloads~\cite{azure_trace}, making cold starts inevitable in serverless LLM serving~\cite{serverlessllm,lambdascale,blitzscale,deepflow}.
Cold starts can directly impact service level objectives (SLOs) for LLM inference.
For instance, in our production environment, a cold-start instance produces the first token after more than 40 seconds, whereas subsequent generation takes only $\sim$30ms per token.
This contrast stresses the need for faster cold starts.

Addressing this problem is non-trivial, as LLM cold starts introduce two unique challenges compared to traditional serverless workloads.
First, LLM applications require massive external state,
with model weights fetched on demand from remote registries during cold starts.
Since serverless instances typically operate with constrained network bandwidth,
model fetching becomes the primary bottleneck in cold start latency.
Second, the LLM runtime is highly complex, involving multiple specialized libraries including CUDA runtime~\cite{cudnn}, AI frameworks~\cite{pytorch, olston2017tensorflow}, and inference engines~\cite{tensorRT, vllm, sglang}.
The complex dependencies of these libraries lead to substantial container creation and library loading overhead~\cite{socc24-prewarm}.
In public clouds, applications run in isolated containers with varying library versions, making it impractical to prepare runtimes in advance for every user.

Prior works focus on reducing the model fetching latency and propose two approaches: (1) caching LLMs in memory or SSDs~\cite{serverlessllm, deepflow, faaswap} and (2) retrieving model weights from active workers holding the same model~\cite{blitzscale,lambdascale, deepflow}.
While effective for frequently accessed models, these optimizations fall short for long-tail customers,
who benefit the most from the serverless paradigm~\cite{serverless_survey}.
Additionally, retrieving data from peer workers requires high-bandwidth networking,
which is not cost-efficient for serverless inference clusters.

To this end, we present \sysname, a serverless LLM serving system designed to minimize cold start latency in public clouds.
\sysname incorporates two key insights to mitigate the overhead of model fetching and runtime preparation.
First, while a single server has limited bandwidth for model fetching, distributing workers across multiple servers enables efficient bandwidth aggregation.
Second, since model fetching and runtime preparation exhibit weak execution dependencies,
these stages can be parallelized rather than executed sequentially, thereby overlapping their latencies.

Specifically, to distribute models across servers, we take advantage of the layered structure of LLMs.
\sysname partitions LLM layers across servers and exchanges intermediate results during inference.
This design, known as \emph{pipeline parallelism}, has been well studied in distributed model training and inference~\cite{huang2019gpipe, terapipe, narayanan2019pipedream, nips18_pipesgd, nips22_sapipe, megascale, li2023alpaserve, hpipe, pipeinfer, pipeedge}.
Pipeline parallelism is also used to quickly start inference when scaling up multiple workers~\cite{lambdascale}.
\sysname further adopts pipeline parallelism to reduce the blocking effect of model fetching---we proactively create a group of workers in parallel even if the model only needs one.

Although pipeline parallelism can effectively reduce the cold start latency for the initial inference,
it may impact performance in subsequent rounds due to GPU sharing.
To address this, we propose \emph{pipeline consolidation}.
Specifically, we allow workers to fetch and load the layers previously assigned to other workers in the background, eventually transitioning back to local inference with a fully-loaded model.
Depending on the load, each worker can independently decide whether to release its resources or load the full model after cold starts.
Therefore, with a single cold start, \sysname can choose to only create a single worker in the end or scale up as many workers as the pipeline size, allowing for high elasticity.

\sysname takes a three-level \emph{hierarchical} approach to realize the above optimizations.
At the \emph{cluster level}, we design an algorithm to allocate resources for cold-start models based on the model size, cluster resource usage, and user SLOs.
A network-contention-aware worker placement policy is adopted to avoid potential network contention.
At the \emph{worker level}, we overlap remote-to-host model fetching, host-to-GPU model loading, and the initialization of container and GPU runtime to further reduce the cold start latency.
Finally, at the \emph{inference level}, we consolidate workers that are previously created in pipeline groups and migrate their runtime states to ensure peak performance with all weights loaded.

In summary, we make the following contributions:
\begin{itemize}[leftmargin=*]
    \item We articulate the main bottleneck of LLM cold starts in public clouds and identify opportunities to mitigate the overhead.
    \item We propose a hierarchical approach that combines cluster-level, worker-level, and inference-level optimizations to reduce cold start latency for serverless LLM serving in public clouds.
    \item We implement \sysname in both testbed and production environment. Experiments on real-world datasets show that \sysname reduces the cold start latency by 1.7$\times$-- 4.7$\times$ and improves SLO attainment by 1.43$\times$--1.74$\times$ compared to baselines.
    The brownfield evaluation further confirms these gains, yielding an average latency reduction of 2.6$\times$.
\end{itemize}
 \section{Background and Motivation}
\label{sec:background}

This section provides an overview of large language models (LLMs) and serverless LLM serving.
We highlight the critical issue of limited network bandwidth for LLM serving in public clouds,
which can substantially increase cold start latency.
We then identify opportunities for mitigating this performance bottleneck in a cost-effective manner.

\subsection{LLM Inference}

An LLM takes as input a sequence of tokens, called \emph{prompt}, and generates an output sequence in an \emph{autoregressive} manner,
producing one token at a time based on both the prompt and previously generated outputs.
The core component of an LLM is the self-attention layer, which needs to compute the key, value, and query vectors of tokens in the input sequence.
Since the key and value vectors of previous tokens remain unchanged during iterations,
LLM serving systems usually cache these vectors to avoid redundant computation, known as \emph{KV cache}.
The stage that generates the first token is called \emph{prefill}, during which the key and value vectors of all tokens in the prompt are calculated and stored in GPU memory.
Subsequent tokens are generated in the \emph{decoding} phase, which reuses the key-value cache and generates one token at a time.

In LLM serving, users often specify service level objectives (SLOs) that represent performance expectations~\cite{distserve, llumnix, fastserve,li2023alpaserve}.
These SLOs typically focus on two metrics: time to first token (TTFT) and time per output token (TPOT).
TTFT measures the latency from request submission to the generation of the \emph{first} token,
while TPOT represents the average time taken to generate each \emph{subsequent} token.
The relative importance of these metrics varies across different applications.
For example, real-time chatbots emphasize low TTFT, whereas article writing prefers lower TPOT~\cite{distserve}.

Pipeline parallelism is a model parallelism strategy that is commonly used in LLM training to scale up involved GPUs~\cite{huang2019gpipe, terapipe, narayanan2019pipedream, nips18_pipesgd, nips22_sapipe, megascale, graphpipe}.
This approach distributes a model's layers across multiple workers,
with intermediate results transmitted sequentially between workers during computation.

\subsection{Serverless LLM Serving in Public Clouds}

\begin{table}[t]
    \centering
    \resizebox{\linewidth}{!} {
    \begin{tabular}{cccccc}
        \toprule
        \textbf{Instance}  & \textbf{Mem.(GB)} & \textbf{Band.(Gbps)} & \textbf{\#GPU} & \textbf{Cost(\$/h)} & \textbf{Cost/GPU(\$/h)} \\
        \midrule
        g6e.xlarge & 32 & up to 20 & 1 & 1.861 & 1.861   \\
        g6e.2xlarge & 64 & up to 20 & 1 & 2.24208 & 2.24208   \\
        g6e.4xlarge & 128 & 20 & 1 & 3.00424 & 3.00424   \\
        g6e.8xlarge & 256 & 25 & 1 & 4.52856 & 4.52856   \\
        g6e.16xlarge & 512 & 35 & 1 & 7.57719 & 7.57719   \\
        g6e.12xlarge & 384 & 100 & 4 & 10.49264 & 2.62316   \\
        g6e.24xlarge & 768 & 200 & 4 & 15.06559 & 3.76640   \\
        g6e.48xlarge & 1536 & 400 & 8 & 30.13118 & 3.76640   \\
        \bottomrule
    \end{tabular}
    }
    \vspace{-0.1in}
    \caption{Configurations and costs of L40S instances on AWS EC2. The number of vCPU is proportional to memory capacity.}
    \vspace{-1em}
    \label{tab:aws}
\end{table}

Recently, serverless inference has gained widespread adoption among cloud providers, including Aliyun~\cite{fc-gpu}, AWS~\cite{sagemaker}, and
Azure~\cite{azure-serverless}.
In serverless LLM serving, users upload (1) model weights and (2) an image containing the serving framework and runtime dependencies to a cloud storage.
This serving framework is responsible for fetching model weights from remote storage, loading them into GPU memory, and running inference upon user requests.
The serverless platform automatically scales model workers to accommodate fluctuating loads,
including the ability to scale down to zero workers when idle.
One of the most attractive offerings of serverless LLM serving is its pay-per-use billing, where users are charged only for the running periods of each worker.
In production environments, long-tail models typically exhibit sporadic and unpredictable workload patterns~\cite{muxserve},
making them ideal candidates for serverless serving~\cite{serverlessllm}.

\parabf{Network bandwidth constraints.}
Contrary to conventional wisdom, serverless LLM inference may operate with substantially lower network capacity than commonly anticipated.
This constraint arises partly from multiple instances sharing a single server,
but the primary driver is the cost-efficiency requirement of serverless computing.
LLM inference performance mainly depends on the GPU capabilities.
Since serverless customers prioritize cost savings,
cloud providers aim to minimize \textbf{cost per GPU} when configuring their infrastructure.
This cost-optimization strategy favors servers with reduced CPU, memory, and network resources.

Table~\ref{tab:aws} illustrates this economic principle by comparing configurations and costs of AWS EC2 instance types equipped with NVIDIA L40S GPUs~\cite{aws-price}.
As Table~\ref{tab:aws} reveals,
resources other than GPUs constitute a significant portion of server costs. 
Using single-GPU instances as an example, compared to the instance type with the lowest cost per GPU (g6e.xlarge),
adding extra resources can increase costs by 20\% to 300\%.
Consequently, servers with constrained non-GPU resources deliver substantial cost savings, making them the preferred choice for serverless providers.
However, these economical instances impose network bandwidth limitations that significantly increase cold start latency.
While a network-only upgrade is less expensive,
the additional network bandwidth would be severely underutilized, as it is only used during cold starts for model fetching.
This paper aims to address this trade-off by minimizing cold start latency while preserving the cost benefits of resource-constrained servers.

\begin{figure}[t]
    \centering
    \includegraphics[width=\linewidth]{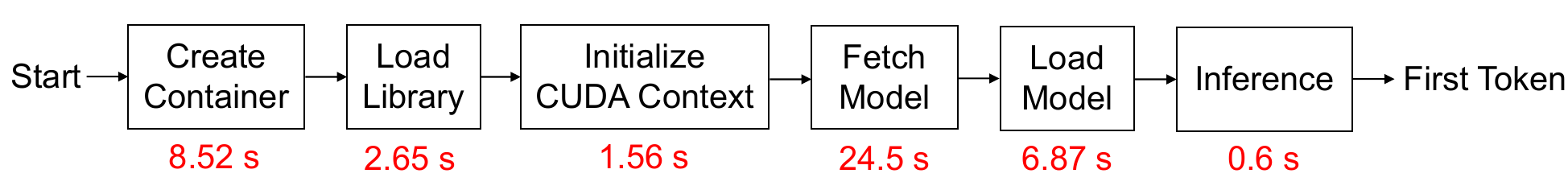}
    \vspace{-0.25in}
    \caption{Cold start latency breakdown.}
    \label{fig:breakdown}
    \vspace{-1em}
\end{figure}

\parabf{Breakdown of LLM cold starts.}
Despite the potential benefits of serverless LLM serving,
existing systems suffer from significant cold start latency.
A \emph{cold start} occurs when an incoming request finds no available worker hosting the target model,
necessitating the creation of a new worker.
Figure~\ref{fig:breakdown} provides a detailed breakdown of cold start latency in our public serverless inference platform~\cite{fc-gpu}, using vLLM~\cite{vllm} to run a Llama2-7B model on NVIDIA A10 GPUs.
In general, a cold start involves the following stages:
\begin{itemize}[leftmargin=*]
    \item \textbf{Container Creation.} The cluster controller allocates resources and creates a container on a GPU server.
    \item \textbf{Library Loading.} The container starts the Python runtime and imports libraries like PyTorch~\cite{pytorch} and TensorFlow~\cite{tensorflow}, along with LLM serving frameworks such as vLLM~\cite{vllm}, SGLang~\cite{sglang}, and TensorRT-LLM~\cite{tensorRT}.
    \item \textbf{CUDA Context Initialization.} The runtime initializes the CUDA context to prepare for GPU tasks.
    \item \textbf{Model Fetching.} The serving framework retrieves the model from remote storage to local memory.
    \item \textbf{Model Loading.} The fetched model is loaded into GPU memory and model states including CUDA graph and KV cache are initialized.
    \item \textbf{Inference.} Run the model and generate the first token.
\end{itemize}

As shown in Figure~\ref{fig:breakdown}, a cold start requires more than 40 seconds to produce the first token.
Model fetching dominates this latency due to limited network bandwidth and contention among colocated containers.
Additionally, container creation incurs substantial overhead due to the large image sizes of LLM workloads (8.31 GB in our setup).
For model loading, prior work has largely eliminated CUDA graph and KV cache initialization overhead through state materialization~\cite{medusa}.
The remaining bottleneck lies in transferring model weights to the GPU, which takes approximately 2 seconds in our setup.

\begin{figure}
    \centering
    \includegraphics[width=\linewidth]{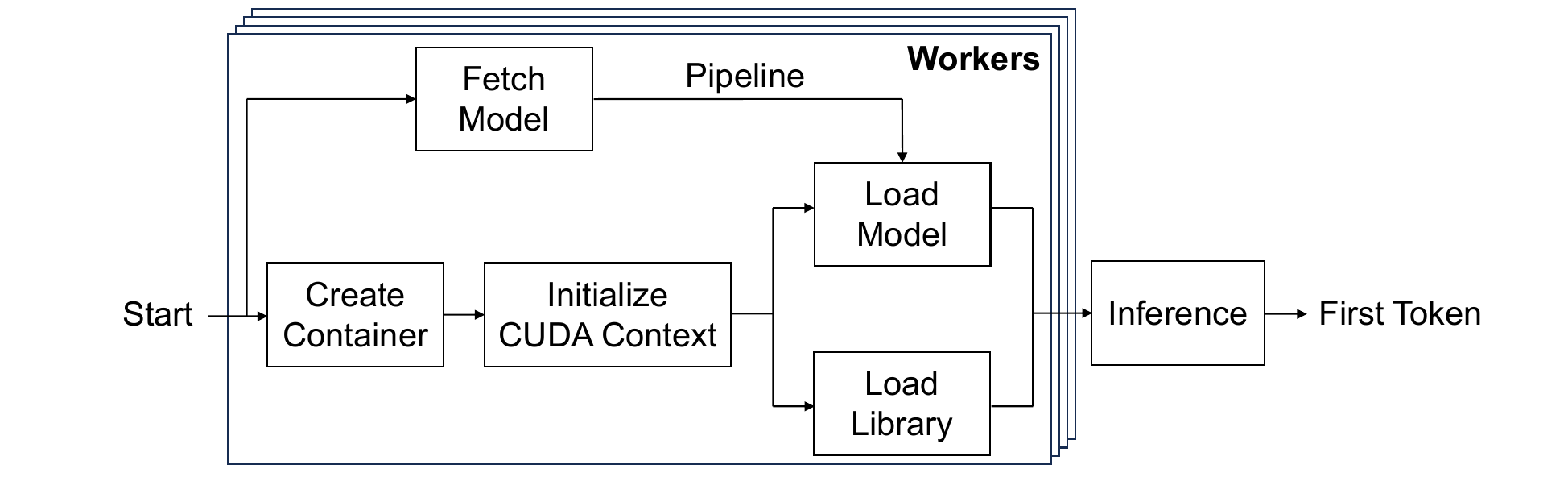}
    \vspace{-0.25in}
    \caption{Optimized cold-start workflow.}
    \label{fig:opt-breakdown}
    \vspace{-1em}
\end{figure}

\subsection{Opportunities}

This paper achieves maximum overlapping to reduce cold start latencies.
At the cluster level, we leverage pipeline parallelism for coarse-grained overlapping of model fetching across workers.
Upon a cold start, we create a pipeline parallelism group across GPU servers, with each worker only hosting a part of the model.
This approach can significantly reduce the single-worker startup latency.
At the worker level, we perform fine-grained overlapping by carefully reorganizing the intra-worker workflow in Figure~\ref{fig:breakdown}.
We offload model fetching to a system-level service,
allowing fetching to begin before container creation.
Furthermore, we observe that model loading (GPU-bound) and library loading (CPU-bound) use different resources,
allowing parallel execution.
We reorder operations to initialize CUDA context first,
and then simultaneously start to load model and library.
Finally, we pipeline the fetching and loading at tensor granularity to hide the loading overhead.
Figure~\ref{fig:opt-breakdown} shows this optimized workflow.
 \section{\sysname Overview}
\label{sec:overview}

\begin{figure}[t]
    \centering
    \includegraphics[width=\linewidth]{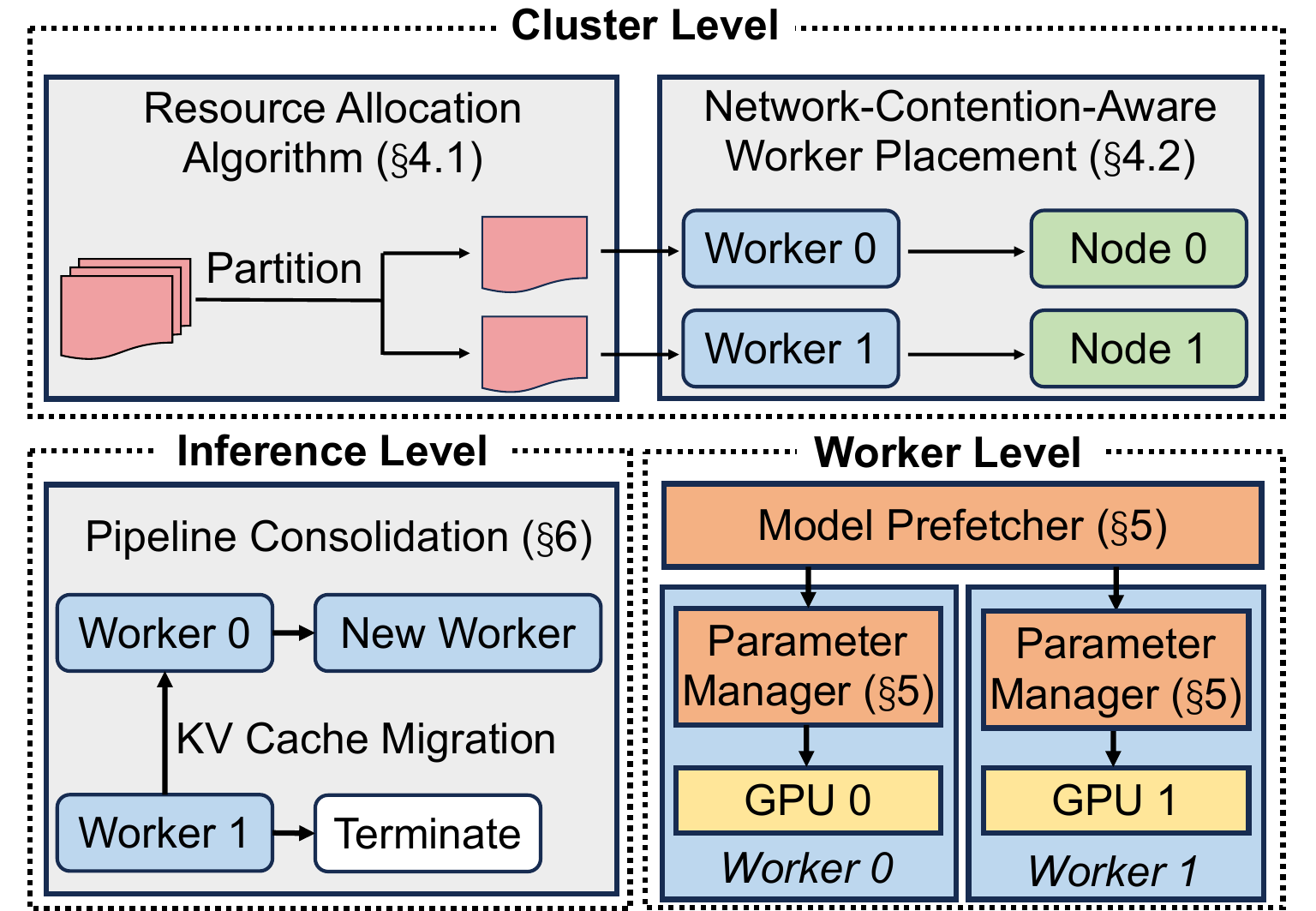}
    \vspace{-0.25in}
    \caption{\sysname system overview.}
    \label{fig:overview}
    \vspace{-1em}
\end{figure}

\sysname is a serverless LLM serving system designed to minimize the cold start latency.
As shown in Figure~\ref{fig:overview}, \sysname adopts a three-level hierarchical architecture.
At the cluster level, the central controller employs a resource allocation algorithm (\S\ref{sec:pipeline}) and a network-contention-aware worker placement strategy (\S\ref{sec:placement}) for parallel model fetching.
The algorithm allocates resources for each cold-start model based on user SLOs,
while the placement strategy identifies potential network contentions on GPU servers and ensures that such contention does not cause SLO violations.
\sysname also strives to distribute pipeline workers across GPUs to mitigate performance degradation caused by GPU sharing.

At the worker level, \sysname applies fine-grained overlapping across cold-start stages to accelerate worker startup (\S\ref{sec:prefetching}).
First, a node-level model prefetcher is used to proactively fetch model weights for cold-start workers, overlapping model fetching with container creation and runtime initialization times.
Second, \sysname prioritizes the CUDA context initialization and introduces a parameter manager that loads parameters to the GPU in parallel with library loading.
Additionally, model fetching and loading are pipelined to hide the model loading overhead.
These overlapping strategies reduce the overhead of cold start stages other than model fetching, thereby highlighting the value of pipeline parallelism.

Finally, at the inference level, \sysname consolidates pipeline workers into standalone workers that host all parameters (\S\ref{sec:elastic}).
Specifically, \sysname allows a cold-start worker to continue loading the remaining model parts while serving requests, eventually evolving into an individual worker that hosts the entire model.
Each worker can independently decide whether to evolve or terminate after completing the cold start process.
Ongoing requests in the pipeline parallelism group are then migrated to the standalone workers.

\sysname ensures that cold starts incur no higher first-token latencies or request completion times than standard cold-start processes.
This guarantee in request completion time is achieved through \sysname's request migration mechanism,
which ensures peak inference speed after all model weights have been prepared.

\begin{figure}[t]
    \centering
    \subfloat[Normal cold start.]
    {\includegraphics[width=0.46\linewidth, page=1]{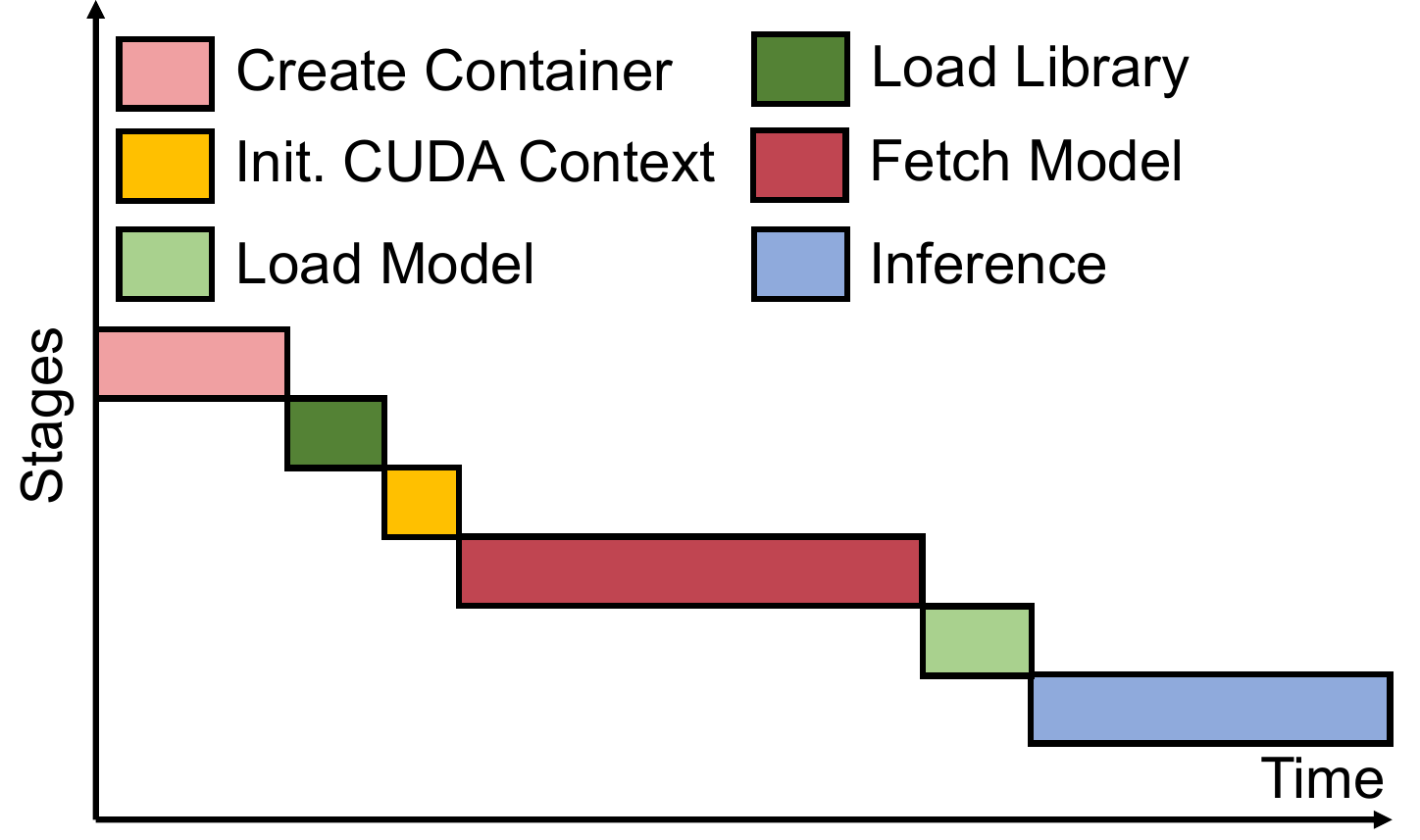}}
    \quad
    \subfloat[Cold start with 2 workers.]
    {\includegraphics[width=0.46\linewidth, page=2]{figures/design/pipeline.pdf}}
    \quad
    \subfloat[Cold start with scaling down.]
    {\includegraphics[width=0.46\linewidth, page=3]{figures/design/pipeline.pdf}}
    \quad
    \subfloat[Cold start with scaling up.]
    {\includegraphics[width=0.46\linewidth, page=4]{figures/design/pipeline.pdf}}
    \quad
    \vspace{-0.1in}
    \caption{Comparison of cold-start workflows for different methods.}
    \label{fig:pipeline}
    \vspace{-1em}
\end{figure}

\begin{figure*}[t]
    \centering
    \subfloat[TTFT of different pipeline parallelism sizes.]
    {\includegraphics[width=0.32\textwidth]{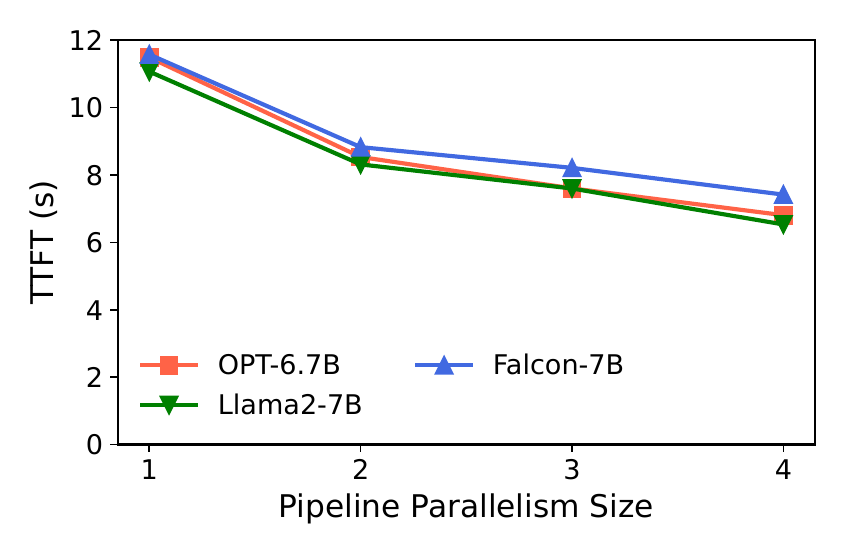}}
    \quad
    \subfloat[Impact of pipeline parallelism size on TPOT.]
    {\includegraphics[width=0.32\textwidth]{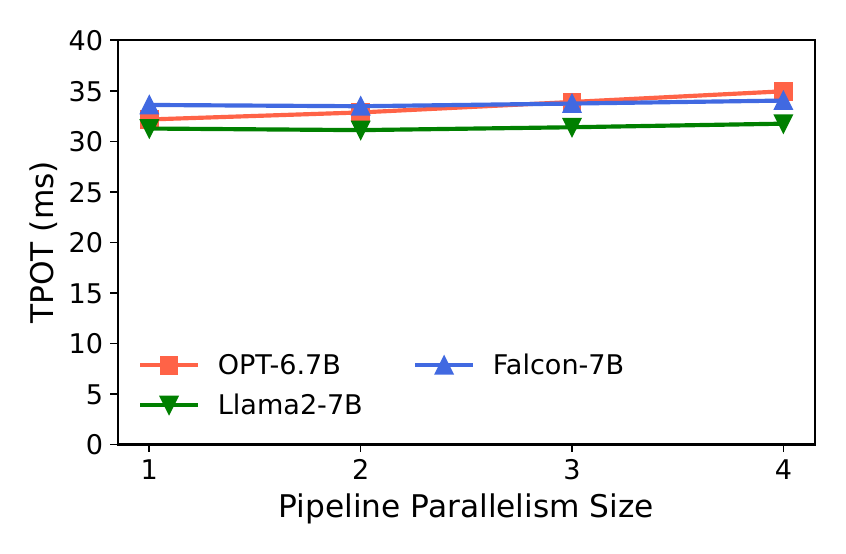}}
    \quad
    \subfloat[Impact of cost on TPOT.]
    {\includegraphics[width=0.32\textwidth]{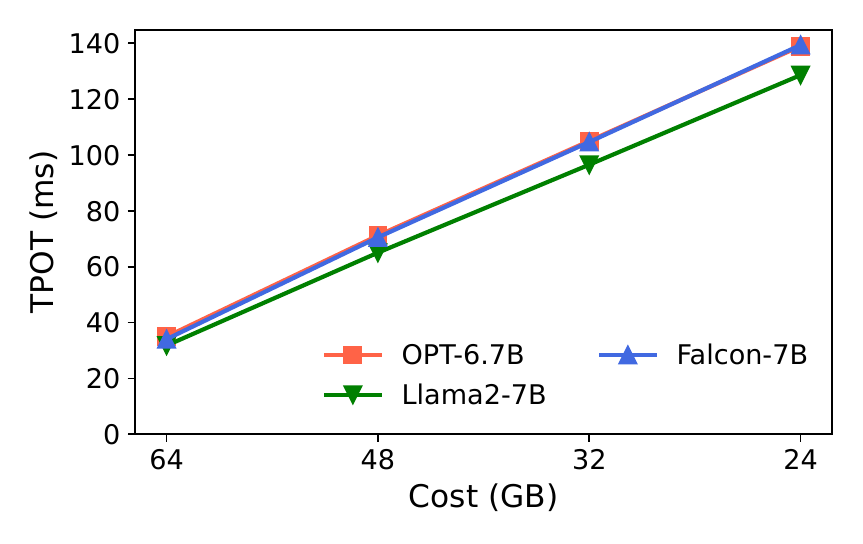}}
    \vspace{-0.1in}
    \caption{Tradeoff analysis of pipeline parallelism.}
    \label{fig:tradeoff}
    \vspace{-1em}
\end{figure*} \section{Cluster-Level Controller}

In this section, we present the design of the cluster-level central controller in \sysname.
Figure~\ref{fig:pipeline}(a) illustrates the normal cold-start workflow, which goes through six stages sequentially.
\sysname leverages pipeline parallelism to optimize model fetching, the most time-consuming stage.
As shown in Figure~\ref{fig:pipeline}(b), upon a cold start, we launch multiple workers on different servers, each fetching a part of the model.
During inference, workers exchange intermediate results but the sizes of these messages are relatively small.
By reducing the portion of the model each worker needs to fetch, pipeline parallelism significantly reduces cold start latency.
\sysname finds the optimal resource allocation scheme for each cold-start model based on the model size and user SLOs (\S\ref{sec:pipeline}),
and employs a network-contention-aware worker placement strategy to prevent SLO violations (\S\ref{sec:placement}).

\subsection{Resource Allocation}
\label{sec:pipeline}

To design an efficient resource allocation algorithm for pipeline workers,
we begin by quantifying the tradeoffs involved in pipeline parallelism.
Based on this analysis, we propose a TTFT and worst-case TPOT prediction method.

\parabf{Tradeoff analysis.}
The benefits of pipeline parallelism come with performance and resource overhead.
To better understand these tradeoffs, we conduct experiments using four GPU servers, each equipped with a NVIDIA A10 GPU and 188 GB memory.
The network bandwidth of each server is 16 Gbps.

Figure~\ref{fig:tradeoff}(a) demonstrates the TTFT of models under different pipeline parallelism sizes.
Larger parallelism sizes reduce model fetching time, resulting in shorter TTFT.
Note that the marginal improvement in TTFT diminishes as other cold-start stages also consume non-negligible time.
This limitation is further addressed by worker-level optimizations in \S\ref{sec:prefetching}.

Pipeline parallelism has a modest impact on TPOT, as shown in Figure~\ref{fig:tradeoff}(b).
This is because the messages transmitted between workers are relatively small.
For example, Llama2-7B incurs only 8 KB of inter-layer results per token.

However, pipeline parallelism may lead to worker colocations across different models.
For a model initially requiring $M$ GPU memory,
distributing it across $S$ GPUs reduces per-GPU memory consumption to $M/S$.
Under heavy load, the cluster must co-place workers whose combined memory requirements fit within a single GPU's capacity to maximize resource utilization.
This colocation of workers from different models can degrade performance.

Figure~\ref{fig:tradeoff}(c) illustrates this issue by changing the GPU memory allocated to each model,
with pipeline parallelism size fixed at 4.
Allocated GPU memory represents the cost of a model,
and lower memory usage per model leads to worker colocation.
For example, allocating 32GB GPU memory to each model makes three models share four GPUs.
As the cost decreases, each GPU hosts multiple concurrently running workers.
Consequently, the GPU's computational resources are allocated proportionally to each worker's reserved memory, resulting in longer TPOT.

In conclusion, while increasing pipeline parallelism size reduces the cold start latency, it may impact inference performance due to GPU sharing.
Allocating additional GPU memory to workers can mitigate this performance loss, but at the expense of higher resource consumption.
\sysname's resource allocation algorithm systematically balances these tradeoffs to optimize system performance.

\parabf{Algorithm design}.
The resource allocation algorithm finds an appropriate resource allocation strategy for each cold-start model according to user SLOs.
Apart from pipeline parallelism size,
the algorithm should decide how to place pipeline workers and allocate how much GPU memory to each worker.

To narrow down the search space, we limit the GPU memory allocation choices for each worker to two cases:
(a) the same as the non-parallelized setup, and (b) the minimal memory required to perform inference (proportional to the inverse of pipeline parallelism size).
We call them full-memory workers and low-memory workers, respectively.

Our algorithm has two stages.
First, we enumerate all deployment choices with different pipeline parallelism sizes or GPU memory usage and predict the TTFT and worst-case TPOT for each choice.
Second, we select the optimal choice that satisfies user SLOs with minimum resource usage.
The maximum pipeline parallelism size is limited to $4$ as larger parallelism sizes yield little improvement.

The TTFT and TPOT prediction takes historical information as the input.
Formally, it includes the time cost of container creation and runtime initialization ($t_c$), data transmission ($t_n$), prefill ($t_p$), and decoding ($t_d$).
The data transmission time is the latency of the TCP network between GPU servers.
The prefill and decoding time costs are model-specific metrics obtained from the model's previous executions.
Assume the pipeline parallelism size for the current deployment choice is $s$ and there are $w$ full-memory workers $(0\leq w\leq s)$,
while the network and PCIe bandwidth of servers on which workers reside are $\{b_{q_1},b_{q_2},\cdots,b_{q_s}\}$ and $\{p_{q_1},p_{q_2},\cdots,p_{q_s}\}$.
Finally, we predict the TTFT using the following equation.
\begin{equation}
\begin{split}
\text{TTFT}&=t_c+\frac{M}{s}\times\max_i\{\frac{1}{b_{q_i}}+\frac{1}{p_{q_i}}\}\\
&+t_p\times(s-w+\frac{w}{s})+t_n\times s,
\end{split}\label{eq:prediction-ttft}
\end{equation}
\noindent where $M$ is the model size.
Eq.~\ref{eq:prediction-ttft} is made up of four parts: runtime initialization, model fetching, model loading, and prefilling.
The model fetching and loading time is determined by the partitioned model size and speed of the slowest worker.
During prefilling,
a full-memory worker costs $t_p/s+t_n$ units of time, while a low-memory worker costs $t_p+t_n$ units.

To satisfy TTFT requirement, we select GPU servers to minimize the model fetching and loading time for a given pipeline parallelism size $s$ and number of full-memory workers $w$.
Assume the GPU servers that can accommodate full-memory workers are $\{i_1,i_2,\cdots,i_k\}$.
Apart from them, there are also servers that can accommodate low-memory workers, denoted as $\{j_1,j_2,\cdots,j_l\}$.
Our selection strategy is to first allocate the top $w$ servers with minimum model fetching and loading time (i.e., the smallest $\frac{1}{b_{i_x}}+\frac{1}{p_{i_x}}$) from $\{i_x\}$ to full-memory workers,
and then merge the remaining servers into the set $\{j_x\}$.
After that, we select the top $s-w$ servers from the merged set by the same strategy and allocate them to all low-memory workers.

Similar to the calculation of prefilling latency, we can predict the TPOT as follows.
\begin{equation}
    \text{TPOT}=t_d\times(s-w+\frac{w}{s})+t_n\times s.
\label{eq:prediction-tpot}
\end{equation}
\indent Among all choices that satisfy both TTFT and TPOT SLOs, \sysname prioritizes free GPUs during worker placement.
When the cluster is not under heavy load, this approach improves the inference performance of models.
Algorithm~\ref{alg:pipeline} outlines the resource allocation algorithm.

\begin{algorithm}[t]
    \caption{Resource Allocation Algorithm}\label{alg:pipeline}
    \textbf{Input:} time cost of container creation and runtime initialization $t_c$, data transmission $t_n$, prefill $t_p$, and decoding $t_d$;
    model size $M$; GPU server network bandwidth $b_i$ and PCIe bandwidth $p_i$; user specified requirements $\text{SLO}_{\text{TTFT}}$ and $\text{SLO}_{\text{TPOT}}$.

    \textbf{Output:} pipeline parallelism size $s$, \#full-memory workers $w$, and selected GPU servers $g$.
    \begin{algorithmic}
        \State $S \gets \emptyset$
        \For{$s\in \{1,2,\cdots,4\}$}
        \For{$w\in \{0,1,\cdots,s\}$}
        \State $i_1,i_2,\cdots,i_k\gets$ Servers that fit a model of size $M$.
        \State $j_1,j_2,\cdots,j_l\gets$ Servers that fit a model of size $M/s$.
        \State $j_1',\cdots,j_{l'}'\gets \text{MergeSort}((j_1,\cdots,j_l), (i_{w+1},\cdots,i_k))$
        \State $g\gets (i_1,i_2\cdots,i_w,j_1',\cdots,j_{s-w}')$
        \State $\text{max\_ratio}\gets \max_{x\in g}\left(\frac{1}{b_x}+\frac{1}{p_x}\right)$
\State $\text{TTFT}\gets t_c+\frac{M}{s}\times \text{max\_ratio}+t_p\times(s-w+\frac{w}{s})+t_n\times s$
        \State $\text{TPOT}\gets t_d\times(s-w+\frac{w}{s})+t_n\times s$
        \If{$\text{TTFT}\leq \text{SLO}_{\text{TTFT}}$ \textbf{and} $\text{TPOT}\leq \text{SLO}_{\text{TPOT}}$}
        \State $S \gets S\cup\{(s,w,g)\}$\EndIf
        \EndFor
        \EndFor
        \If{$S~\text{is}~\emptyset$}
        \State \textbf{return} $(1,1, (i_1))$ \Comment{Use single worker if no solution}
        \Else
        \State $c\gets$ Scheme that incurs minimal GPU sharing from $S$
        \State \textbf{return} $c$
        \EndIf
    \end{algorithmic}
\end{algorithm}

\subsection{Network-Contention-Aware Worker Placement}
\label{sec:placement}

There are two types of network contentions in the cluster.
The first one is the contention between model fetching and inference, i.e., the sending of intermediate results across workers.
Because the intermediate results are small, simply prioritizing inference packets solves the contention.

The second type of contention is the interference among cold-start workers on the same GPU server,
which leads to unpredictable cold start performance.
To solve this problem,
we propose a network-contention-aware worker placement policy that places cold-start workers based on user SLOs.

\parabf{Policy design}.
Initially, colocated workers share the network bandwidth with equal credits.
Once we assign a worker to a GPU server,
we record the model size it needs to fetch and the user-specified maximum TTFT.
To place a new worker,
we inspect each GPU server to check whether adding a cold-start worker to this server would lead to SLO violations for existing workers.
If the check passes, we estimate and record the network bandwidth that the new worker can obtain.
Note that PCIe bandwidth is much higher than network and PCIe switch is able to further isolate PCIe usage across tasks, so we do not take PCIe contention into account.

Specifically, we denote the bandwidth of a GPU server as $B$.
For each cold-start worker $i$ on the GPU server, we record the fetching deadline $D_i$ and estimate its pending model size $S_i$.
The fetching deadline comes from the prediction of TTFT (Eq.~\ref{eq:prediction-ttft}).
When a new cold-start worker comes, the network bandwidth per work decreases,
and we check whether all cold-start workers on the server are able to finish timely under the new bandwidth.
Formally, we check whether the following condition is true for all workers.
\begin{equation}
S_i\leq \frac{B}{N+1}\times(D_i-T),
\label{eq:check}
\end{equation}
\noindent where $N$ represents the number of existing workers on the server and $T$ represents the current time. This server accepts the new worker if all workers passed the check.

To estimate the pending model size, we record the time of last network bandwidth change $T$'.
The start and completion of a cold start change the network bandwidth, informing the controller to adjust the pending model size.
Here, we assume the current time is $T$ and the number of cold-start workers before change is $N$.
Then we adjust the pending model size upon each bandwidth change according to the following equation.
\begin{equation}
    S_i'=S_i-\frac{B}{N}\times(T-T').
\label{eq:estimation}
\end{equation}

$S_i'<0$ means that the worker has fetched the model ideally, thus we delete it from the cold-start worker list.

\section{Worker-Level Overlapping}
\label{sec:prefetching}

As described in \S\ref{sec:pipeline}, the TTFT of pipeline parallelism has diminishing marginal returns due to container initialization stages, including container creation, library loading and CUDA context initialization.
In this section, we present the worker-level overlapping in \sysname that carefully reorganizes the startup workflow to reduce worker initialization time.
Basically, \sysname leverages two optimizations to overlap initialization stages.
First, we prefetch model weights on local GPU server as soon as container creation starts so that the container initialization stages are overlapped with model fetching.
Second, we prioritize CUDA context initialization and develop a parameter manager to load model while initializing Python libraries so that the library loading stage is overlapped with model loading.

\begin{figure}
    \centering
    \subfloat[Prefetching whole model.]
    {\includegraphics[width=0.46\linewidth, page=5]{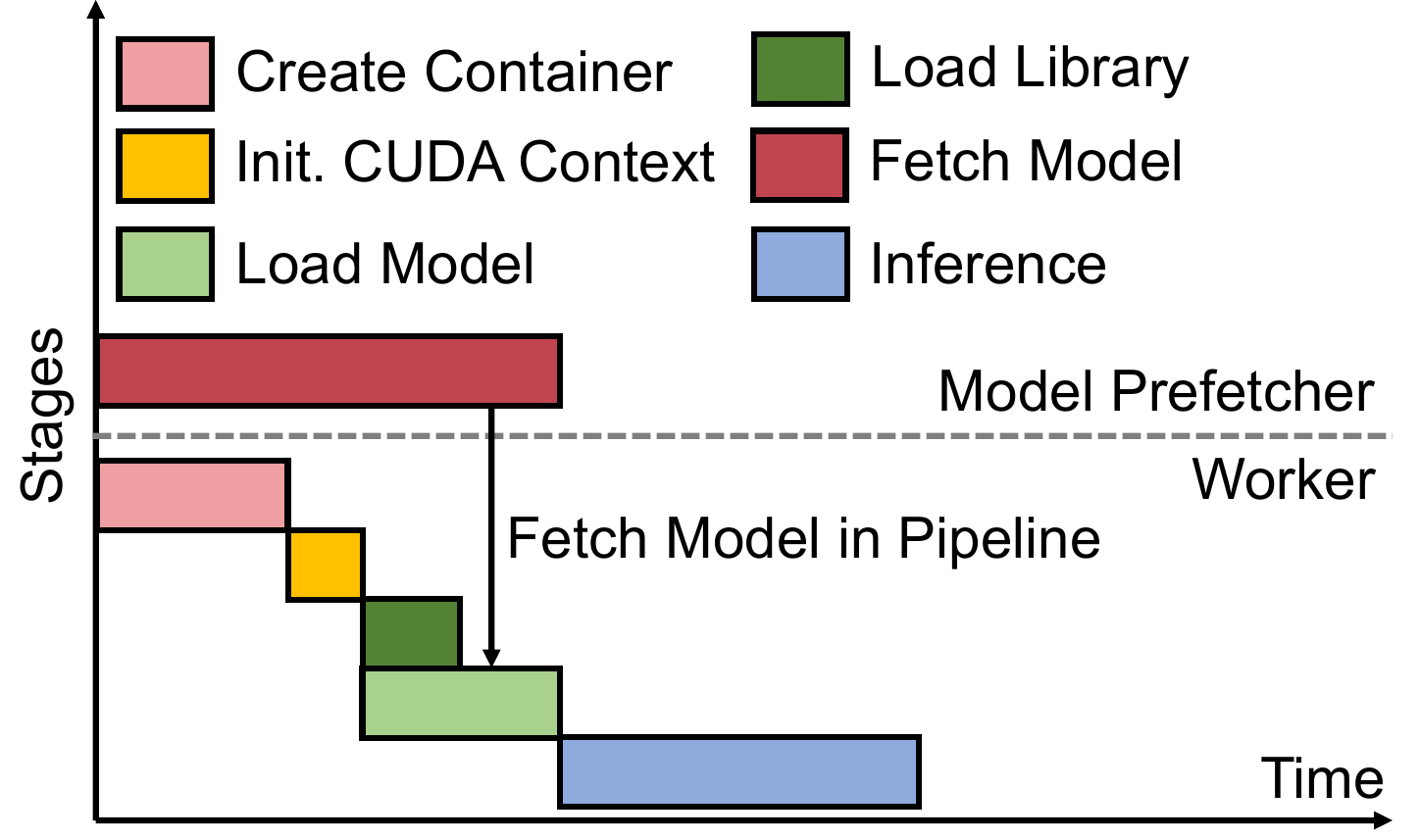}}
    \quad
    \subfloat[Prefetching in two stages.]
    {\includegraphics[width=0.46\linewidth, page=6]{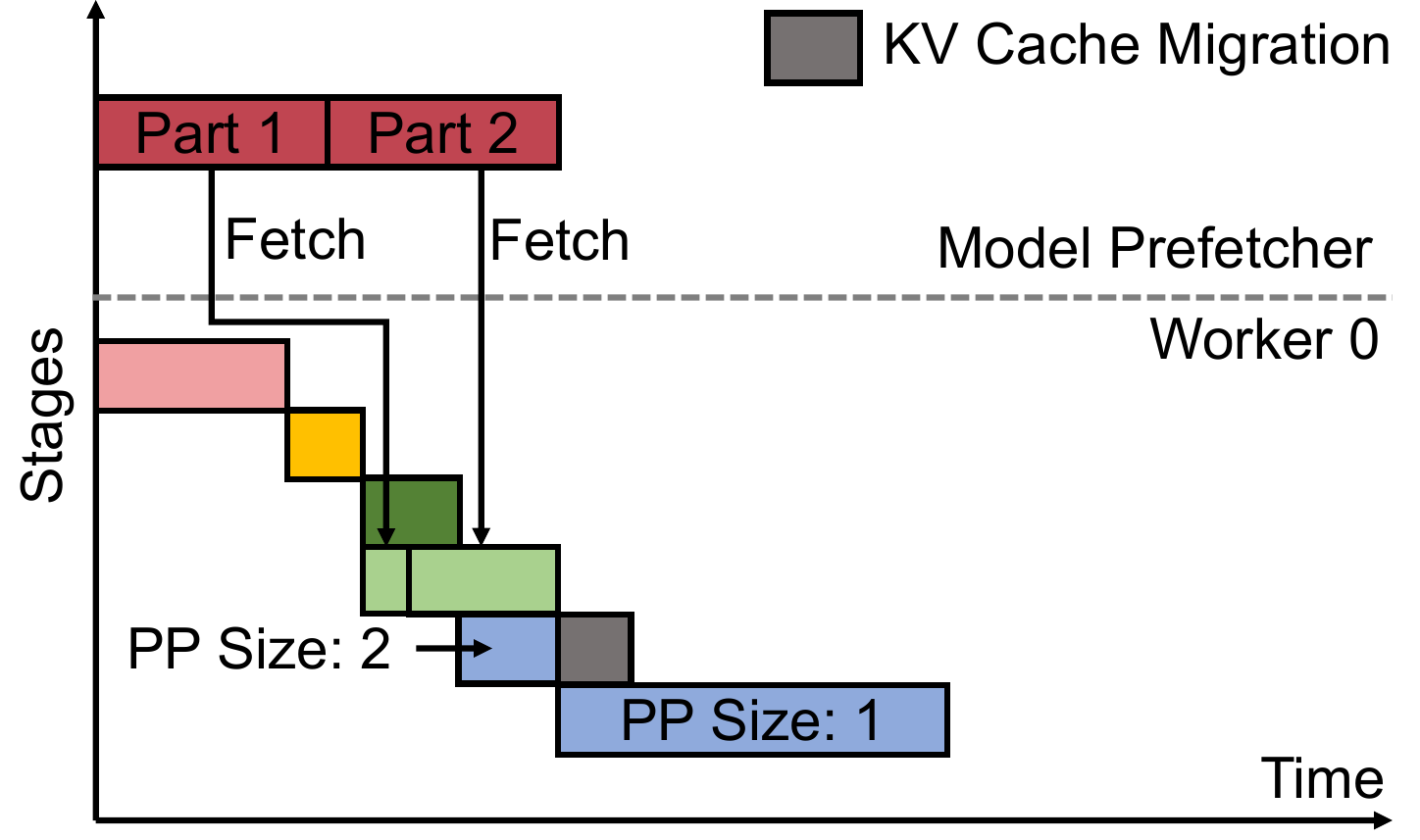}}
    \quad
    \vspace{-0.1in}
    \caption{Cold-start workflows with worker-level overlapping.}
    \label{fig:overlap}
    \vspace{-1em}
\end{figure}

\subsection{Model Prefetching}

\sysname launches a model prefetcher on each GPU server, which is responsible for proactively fetching models from remote storage.
When a worker has been allocated to the server, the central controller informs the model prefetcher about model metadata.
After that, the prefetcher starts to load the model weights from remote storage to a shared memory region.
The cold-start worker will fetch parameters from shared memory in a streaming manner after runtime has initialized.

Model prefetching starts before container creation,
so that the container creation and runtime initialization stages are overlapped.
The worker performs model prefetching and loading in a pipelined fashion.
In the shared memory region of a model,
we use the first eight bytes to store the address that represents the end of currently fetched model weights.
Model weights are represented using the SafeTensors format~\cite{safetensors}.
This format contains the metadata of all parameters at the beginning of the file,
so that it is convenient for the worker to check whether a tensor has been fetched.

As allocating shared memory is time-consuming,
the model prefetcher allocates a shared memory region for all models in advance.
During startup, it accesses each virtual page in the region to allocate corresponding physical pages.
When a fetching request arrives,
the model prefetcher calculates the size of all model files and allocates space from the shared memory region.
A standalone process is then triggered to read the model weights from remote storage and write contents into shared memory.

\subsection{Parameter Manager}

\sysname leverages a parameter manager to load model parameters in the background.
The parameter manager runs in an individual thread and is responsible for resolving tensor metadata, reading weights from the shared memory, and finally loading weights into the GPU.
The whole procedure is zero-copy and pipelined.
The parameter manager also takes advantage of the high parallelism of GPU cores and uses multiple CUDA streams to load models.
The priorities of CUDA streams are determined according to whether the loading process is on the critical path or in the background.

During cold starts, the entry program in container will first initialize the parameter manager to start loading model parameters,
and then import other AI libraries.
The serving framework later queries the parameter manager through a specified API to obtain tensors in a streaming manner with zero copy.

Figure~\ref{fig:overlap}(a) shows the optimized cold-start workflow after adopting model prefetching and parameter manager.
Model fetching and runtime preparation runs in parallel,
while library loading is overlapped with model loading.
In this way, the overhead of other cold start stages is concealed behind model fetching,
reinforcing the effectiveness of pipeline parallelism.

As \sysname may launch workers in pipeline parallelism groups and later load the remaining part of model in background,
Figure~\ref{fig:overlap}(b) also shows the cold-start workflow in this scenario.
The model prefetcher downloads two parts of model sequentially.
After the first part of model has been loaded, the worker starts to perform inference with other workers in the pipeline parallelism group.
Meanwhile, the parameter manager loads the second part of model in background and finally transforms itself into a standalone worker.

\parabf{TTFT prediction with worker-level overlapping.}
After applying worker-level optimizations, the TTFT is reduced by fine-grained overlapping.
Based on additional historical information including the time cost of container creation ($t_{cc}$), CUDA context initialization ($t_{cu}$) and library loading ($t_l$),
we change the TTFT prediction used in resource allocation (Eq.~\ref{eq:prediction-ttft}) into the following formula.
\begin{equation}
\begin{split}
    \text{TTFT}&=\max_i\left(t_{cc}+t_{cu}+\max\left(\frac{M/s}{p_{q_i}}, t_l\right),\frac{M/s}{b_{q_i}}\right)\\
    &+t_p\times(s-w+\frac{w}{s})+t_n\times s.
\end{split}\label{eq:prediction-ttft-opt}
\end{equation}

\section{Inference-Level Consolidation}
\label{sec:elastic}

As \sysname creates a group of workers in parallel when the model only needs one,
we propose pipeline consolidation that merges pipeline parallelism groups into individual workers to obtain optimal performance.
Specifically, after the parameter manager has loaded the requested model parameters in pipeline-parallel inference,
we inform it to continue loading the remaining part of model in background.
Once loading completes, the worker returns to the non-parallelized setup and serves subsequent requests with the whole model in local GPU, thereby achieving optimal performance.
The second part of model is loaded in low-priority CUDA streams,
so that the performance of inference task will not be affected.

\subsection{Worker Scaling}

\sysname provides two scaling choices.
First, after workers have loaded corresponding parts of models,
we can perform \emph{scaling down} by allowing only one of them to fetch the unloaded model parts in background.
Once all parameters are loaded, we migrate all existing requests to that worker along with their key-value cache.
Finally, the worker continues to generate tokens with whole model while other workers are terminated.
Figure~\ref{fig:pipeline}(c) demonstrates the process of scaling down.
In this way, we produce the first token earlier by parallelized model fetching and finally obtain a worker with whole model, similar to standard cold starts.

The second scaling choice is \emph{scaling up}, converting all cold-start workers into individual serving endpoints, as illustrated in Figure~\ref{fig:pipeline}(d).
This approach tackles load spikes,
where the autoscaler aims to deploy multiple workers for the model simultaneously.
By enabling the rapid creation of workers within pipeline parallelism groups,
this design allows the system to achieve maximum throughput more quickly.

By default, \sysname adopts the scaling down mechanism to reduce model fetching latency with minimal overhead.
To promptly react to bursty loads and transition to the scaling up mechanism, we manage worker lifecycles with a sliding window strategy.
For each model, the number of requests received in the previous window is recorded and used to predict the maximum number of requests likely to arrive in the next window.
The required number of workers is then determined based on the current waiting queue length combined with the predicted maximum number of requests expected to arrive in the next window.
In cold start, we create a pipeline parallelism group that is no smaller than required number of new workers and later transform the group into desired number of individual workers.
To handle sudden request surges, multiple pipeline parallelism groups can be created as needed.

\subsection{Key-Value Cache Migration}

After reducing the pipeline parallelism size,
we migrate all uncompleted requests,
allowing for earlier release of resources occupied by other workers.
Since model layers are assigned to different workers, the KV cache of these requests is distributed across workers,
necessitating the KV cache migration.

\sysname performs KV cache migration inside a pipeline parallelism group, where layers are distributed across workers.
Before migration, we first stop scheduling of existing requests and wait for all on-the-fly batches return.
Next, we query the cache block manager to obtain the blocks that are used by existing requests,
and then collect these blocks from all workers with a \emph{gather} operation.
Blocks are gathered to the worker with whole model and placed at different layers, according to which worker it comes from.

The KV cache migration process in \sysname is highly optimized.
First, the whole migration workflow is performed in an individual thread and uses low-priority CUDA streams so that the inference tasks will not be affected.
Second, we create multiple CUDA streams when moving data from or to GPUs to utilize the GPUs' high parallelism.
Furthermore, the data transmission is performed in a streaming manner.
On the target worker, once a chunk of tensors arrives,
it is instantly loaded to GPU in a separate CUDA stream.
On other workers, once a chunk of tensors is loaded from GPU to host,
we immediately send them to the network.

 \section{Implementation}
\label{sec:implementation}

We implement \sysname in both testbed environments and our public serverless inference platform.
Our implementation is based on vLLM~\cite{vllm},
with a modification of around 3200 lines of code in C++ and Python.
For testbed deployment, we develop a custom serverless LLM serving framework with around 3000 lines of code in Python.
This framework makes scheduling decisions and automatically scales number of vLLM workers for each model to match current loads.
The code of \sysname is open-source and is publicly available at {\urlstyle{same}\url{https://github.com/LLMServe/hydraserve}}.

\parabf{Instance startup optimizations.}
We perform several optimizations to the initialization process of vLLM.
First, we postpone the allocation of key-value cache swapping space on CPU to reduce cold start latency.
Second, vLLM runs a profiling forward to obtain the amount of free memory during inference.
Unlike prior work that profiles this information offline~\cite{medusa}, we directly calculate the available free memory based on sizes of intermediate results to skip the online profiling phase.
Third, as vLLM first initializes the model on CPU,
we allow the instance to directly use GPU tensors provided by the parameter manager through overriding tensor metadata.

\parabf{Support for large models.}
\sysname is applicable to both models that reside in a single GPU as well as those deployed in multiple servers with tensor and pipeline parallelism.
Similarly, to improve the cold-start performance of models that are distributed across GPUs,
we can increase their pipeline parallelism size and implement worker-level overlapping.

\begin{figure*}[t!]
    \centering
    \includegraphics[width=0.98\textwidth]{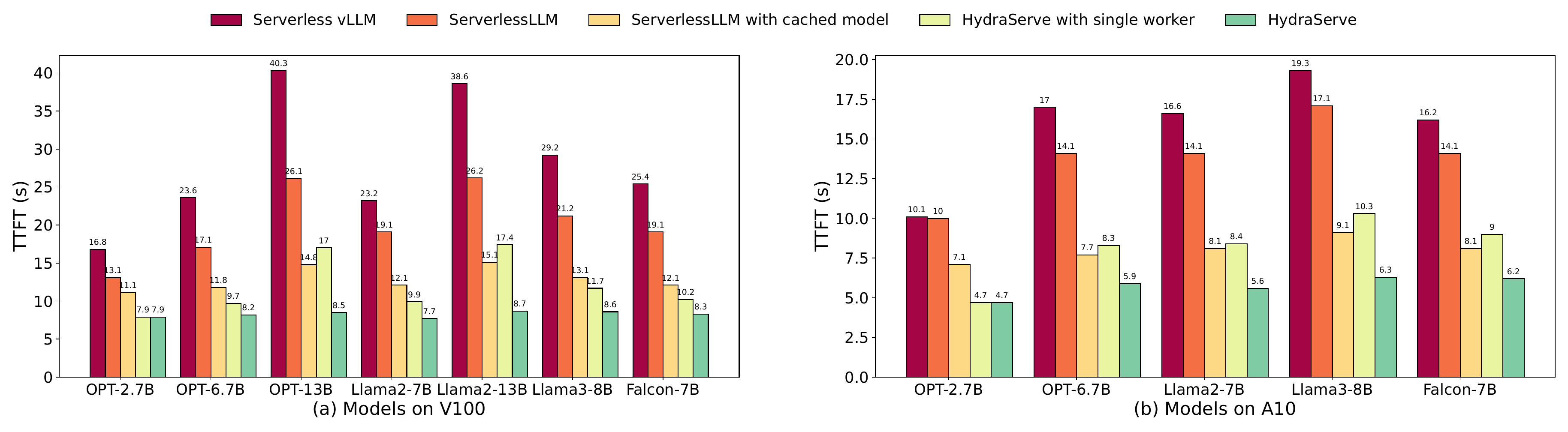}
    \vspace{-0.15in}
    \caption{Cold start latency of systems for different models.}
    \label{fig:eval-ttft}
    \vspace{-1em}
\end{figure*}

\begin{figure}[t]
    \centering
    \includegraphics[width=0.98\linewidth]{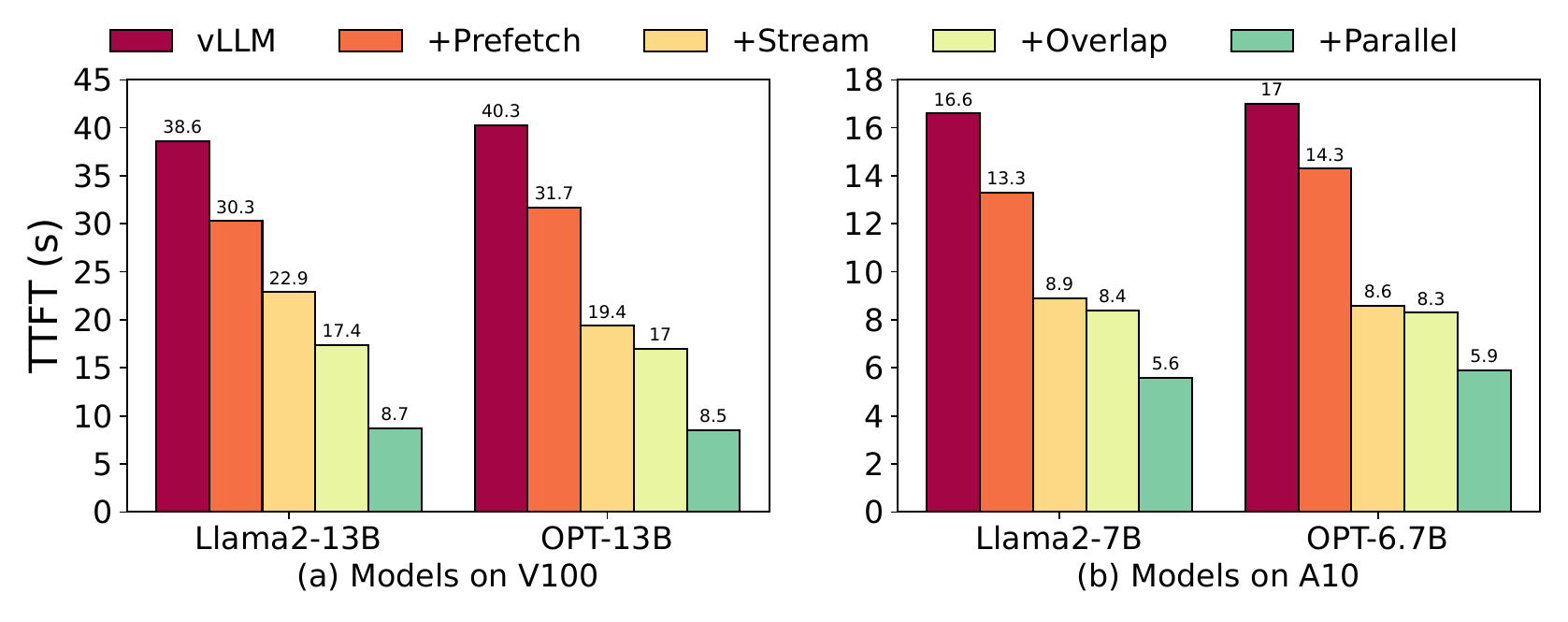}
    \vspace{-0.15in}
    \caption{Performance breakdown of techniques in \sysname.}
    \label{fig:eval-breakdown}
    \vspace{-1em}
\end{figure} \section{Evaluation}
\label{sec:evaluation}

In this section, we present experimental results to validate the efficiency and effectiveness of \sysname.
Our evaluation shows that \sysname achieves a substantial reduction in TTFT, outperforming baselines by 1.7$\times$-- 4.7$\times$ (\S\ref{sec:eval-raw-latency}).
In end-to-end experiments, \sysname improves TTFT SLO attainment by 1.43$\times$--1.74$\times$ across various loads and constraints while maintaining minimal TPOT and cost penalties (\S\ref{sec:eval-end-to-end}).
Our analysis of pipeline consolidation reveals that scaling down reduces end-to-end generation time by up to 2.67$\times$, while scaling up shortens the average TTFT under bursty requests by 1.87$\times$ (\S\ref{sec:eval-consolidation}).
Lastly, we wrap up with brownfield results and show that \sysname achieves an average TTFT reduction of 2.6$\times$ in production environment (\S\ref{sec:deployment}).

\subsection{Experiments Setup}
\label{sec:setup}

\parabf{Testbed.}
We have two testbeds. (i) A GPU cluster that contains 4 A10 servers and 4 V100 servers.
Each A10 server has a single NVIDIA A10 GPU and 188 GB memory, while each V100 server has four NVIDIA V100 GPUs and 368 GB memory.
The network bandwidth per server is 16 Gbps.
(ii) A GPU cluster consisting of 2 servers, each having four NVIDIA A10 GPUs, and another 4 servers, each having four NVIDIA V100 GPUs.
Each A10 server has 752 GB memory and a network bandwidth of 64 Gbps,
while each V100 server has 368 GB memory and 16 Gbps bandwidth.
Both testbeds are connected to a remote model storage that has sufficient network capacity.
Due to the lack of NVLink~\cite{nvlink}, all evaluated models can reside in a single GPU to avoid performance degradation.

\parabf{Baselines.}
We compare \sysname against two baselines.

$\bullet$
Serverless vLLM. vLLM~\cite{vllm} is a LLM serving engine that can serve single model. To serve multiple models, we equip it with the serverless LLM serving framework of \sysname.
During cold starts, the scheduler iterates through all GPU servers and selects the one with sufficient GPU resources to create a new vLLM serving endpoint.

$\bullet$
ServerlessLLM.
ServerlessLLM~\cite{serverlessllm} is the state-of-the-art serverless LLM serving system that reduces the cold start latency by loading-optimized checkpoint and caching.
Due to the lack of high-speed SSDs in our testbeds, we allocate all available server memory for model caching.
We deploy ServerlessLLM on Kubernetes~\cite{kubernetes} and pre-create its containers to eliminate container creation overhead during serving.
The system also uses vLLM~\cite{vllm} as the backend.

We exclude systems that accelerate scaling up via worker cooperation~\cite{lambdascale,blitzscale} from baselines because
(1) they are closed-source and (2) peer fetching offers no benefit over remote storage in our environment.

\begin{figure*}[tp]
    \centering
    \includegraphics[width=0.99\textwidth]{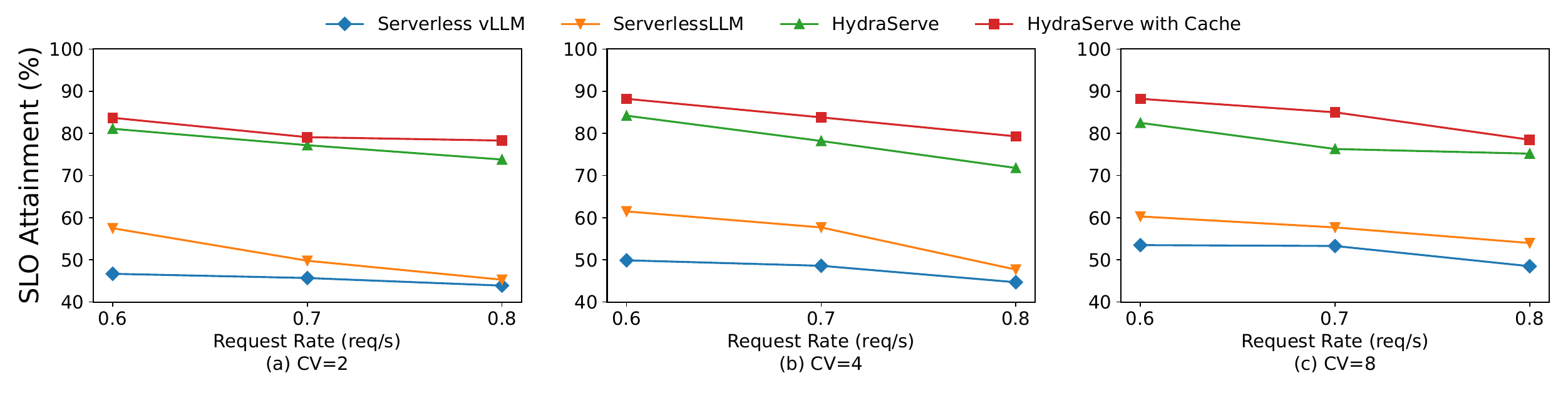}
    \vspace{-0.2in}
    \caption{TTFT SLO attainment of systems under different CVs.}
    \label{fig:eval-end2end-cv}
    \vspace{-1.5em}
\end{figure*}

\begin{figure}[t]
    \centering
    \includegraphics[width=0.99\linewidth]{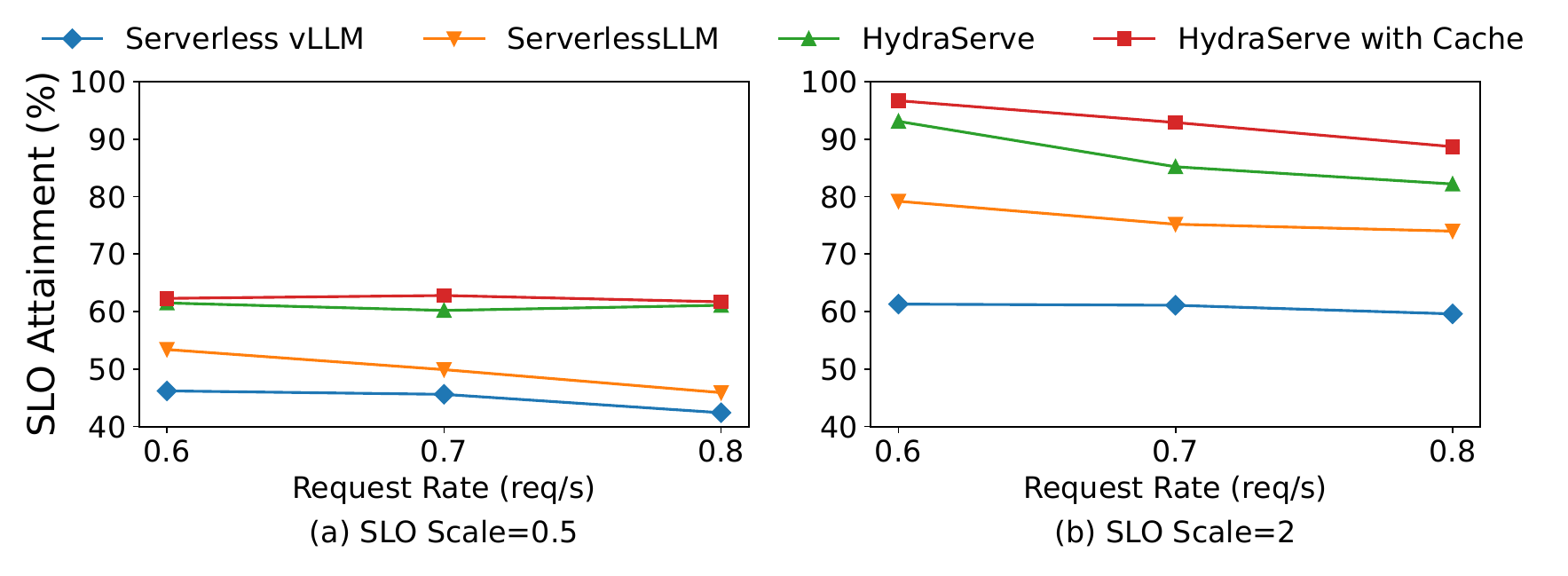}
    \vspace{-0.1in}
    \caption{TTFT SLO attainment of systems under different SLOs.}
    \label{fig:eval-end2end-slo}
    \vspace{-1em}
\end{figure}

\begin{table}[t]
    \centering
    \resizebox{\linewidth}{!} {
    \begin{tabular}{ccccc}
        \toprule
        \textbf{Model}  & \textbf{Model Size} & \textbf{GPU Card} & \textbf{TTFT} & \textbf{TPOT} \\
\midrule
        Llama2-7B & 12.5GB & A10 & 1.5s & 42ms  \\
        Llama2-13B & 24.2GB & V100 & 2.4s & 58ms  \\
        \bottomrule
    \end{tabular}
    }
    \vspace{-0.1in}
    \caption{Measured TTFT and TPOT of warm requests.}
    \vspace{-0.5em}
    \label{tab:slos}
\end{table}

\begin{table}[t]
    \centering
    \resizebox{\linewidth}{!} {
    \begin{tabular}{cccccc}
        \toprule
        \textbf{Application} & \textbf{TTFT} & \textbf{TPOT} & \textbf{Dataset} \\
\midrule
        Chatbot Llama2-7B  & 7.5s & 200ms  & ShareGPT\cite{sharegpt} \\
        Chatbot Llama2-13B & 12s & 200ms & ShareGPT\cite{sharegpt} \\
        Code Completion Llama2-7B & 7.5s & 84ms  & HumanEval\cite{humaneval} \\
        Code Completion Llama2-13B & 12s & 116ms  & HumanEval\cite{humaneval} \\
        Summarization Llama2-7B & 15s & 84ms & LongBench\cite{longbench} \\
        Summarization Llama2-13B & 24s & 116ms & LongBench\cite{longbench} \\
        \bottomrule
    \end{tabular}
    }
    \vspace{-0.1in}
    \caption{Summary of applications in end-to-end experiments.}
    \label{tab:tasks}
    \vspace{-1em}
\end{table}

\subsection{Cold Start Latency}
\label{sec:eval-raw-latency}

We evaluate the cold start latency for systems on testbed (i),
while \sysname is configured at a parallelism size of 4.
For ServerlessLLM, we measure its performance with and without model caching.
Figure~\ref{fig:eval-ttft} illustrates the TTFT of systems for different models,
showing that \sysname achieves the shortest cold start latency across all models.
Specifically, \sysname reduces cold start latency by 2.1$\times$-- 4.7$\times$ compared to serverless vLLM and 1.7$\times$--3.1$\times$ compared to ServerlessLLM.
This improvement is attributed to parallelized model fetching and our worker-level optimizations.

Even with a single worker, \sysname also performs better than ServerlessLLM.
For smaller models like OPT-6.7B,
\sysname with single worker even achieves shorter cold start latency than ServerlessLLM with cached models.
This advantage comes from \sysname's worker-level overlapping strategies and optimizations to vLLM's cold-start process.

\parabf{Performance breakdown.}
To better comprehend the performance of \sysname, we conduct a detailed breakdown of techniques employed in \sysname.
Figure~\ref{fig:eval-breakdown} shows the incremental improvement achieved by applying each proposed technique. 
Starting with the original vLLM system, we apply the following techniques step-by-step: model prefetching (+Prefetch),
streaming loading and implementation optimizations (+Stream), overlapped model and library loading (+Overlap),
and parallelized model fetching (+Parallel).
The results show that each technique contributes to a reduction in cold start latency and the cumulative effect of all techniques results in a substantial overall improvement.

\subsection{End-to-End Experiments}
\label{sec:eval-end-to-end}

We further evaluate the effectiveness of \sysname through comprehensive end-to-end experiments.

\parabf{Workloads.}
We choose the Llama2 model series~\cite{llama2} with FP16 precision in the end-to-end experiments.
Following prior work~\cite{distserve},
we use three typical LLM applications in experiments: chatbot, code completion, and summarization.
Requests for these applications are from ShareGPT~\cite{sharegpt}, HumanEval~\cite{humaneval} and Longbench~\cite{longbench}, respectively.
Since there is no available SLO settings for these applications,
we derive SLOs based on the performance of warm requests.
Specifically, we first measure the TTFT and TPOT for warm requests,
with each request containing 1024 input tokens and a batch size of 8.
The results are shown in Table~\ref{tab:slos}.
We then set the global TTFT SLO to five times the TTFT of warm requests, while applying a stricter TPOT SLO at twice the TPOT of warm requests.
This TTFT SLO is already quite stringent for serverless LLM serving,
considering industrial LLM serving platforms operate with TTFT SLOs as high as 30s~\cite{mooncake}.

Given the nature of summarization tasks, which typically allow more relaxed latency requirements,
their TTFT SLOs are doubled.
Additionally, for chatbot tasks, the TPOT SLO is aligned with standard human reading speeds (i.e., 300 words per minute).
Finally, we generate 64 instances for each application to represent various user models, similar to prior work~\cite{li2023alpaserve, serverlessllm}.
Table~\ref{tab:tasks} shows the summary of applications.

Following prior work~\cite{li2023alpaserve, serverlessllm},
we leverage Microsoft Azure Function Trace~\cite{azure_trace} to generate workloads.
Models are mapped to functions in the trace using a round-robin approach,
and requests are sampled from the trace with Gamma distribution.
We control the sampling process by changing coefficient of variance (CV) and requests per second (RPS).

\parabf{Effectiveness under different CVs.}
Figure~\ref{fig:eval-end2end-cv} demonstrates the TTFT SLO attainment of systems under different CVs.
To assess the effectiveness of cache, we also evaluate \sysname with caching enabled.
The results indicate that as RPS increases, TTFT SLO attainment decreases due to a lack of resources during bursty requests.
Nevertheless, \sysname consistently satisfies most of TTFT SLO requirements under different loads, achieving 1.43$\times$--1.74$\times$ higher TTFT SLO attainment compared to baselines in all scenarios.
This is because \sysname selects appropriate pipeline parallelism size based on user SLOs and distributes cold-start workers to mitigate network contention.
Although some workers are allocated with additional resources to reduce TPOT,
\sysname promptly reclaims these resources after pipeline consolidation to maintain performance under heavy loads.
In contrast, ServerlessLLM exhibits high SLO violations due to limited effectiveness of caching for long-tail models.
Enabling caching in \sysname further improves the TTFT SLO attainment by up to 1.11$\times$ due to benefits for hot models.

For TPOT SLO attainment, both \sysname and baselines achieve over 95\% attainment in most scenarios,
and more than 90\% attainment under all CVs and RPS configurations.
This level of performance is sufficient for most use cases.
Detailed results are provided in the appendix due to space limitations.

\parabf{Effectiveness under different SLO scales.}
We evaluate the systems under different TTFT and TPOT SLOs by adjusting a global SLO scaling parameter, with CV fixed at 8.
Figure~\ref{fig:eval-end2end-slo}(a) shows that, under tight SLOs,
all systems experience significant SLO violations since the time granted for preparing a cold-start worker is exceedingly limited. 
In such cases, the TTFT SLO attainment of all systems is capped at 63\%.
However, \sysname still outperforms baselines, as its faster worker initialization reduces the waiting time for subsequent requests, even if the first request violates SLO.
Figure~\ref{fig:eval-end2end-slo}(b) demonstrates the performance under looser SLO conditions.
\sysname achieves 1.38$\times$--1.52$\times$ improvement in TTFT SLO attainment compared to baselines,
with caching further enhancing this improvement to 1.49$\times$--1.58$\times$.

\begin{figure}[t]
    \centering
    \begin{minipage}[t]{0.49\linewidth}
        \centering
        \includegraphics[width=0.99\linewidth]{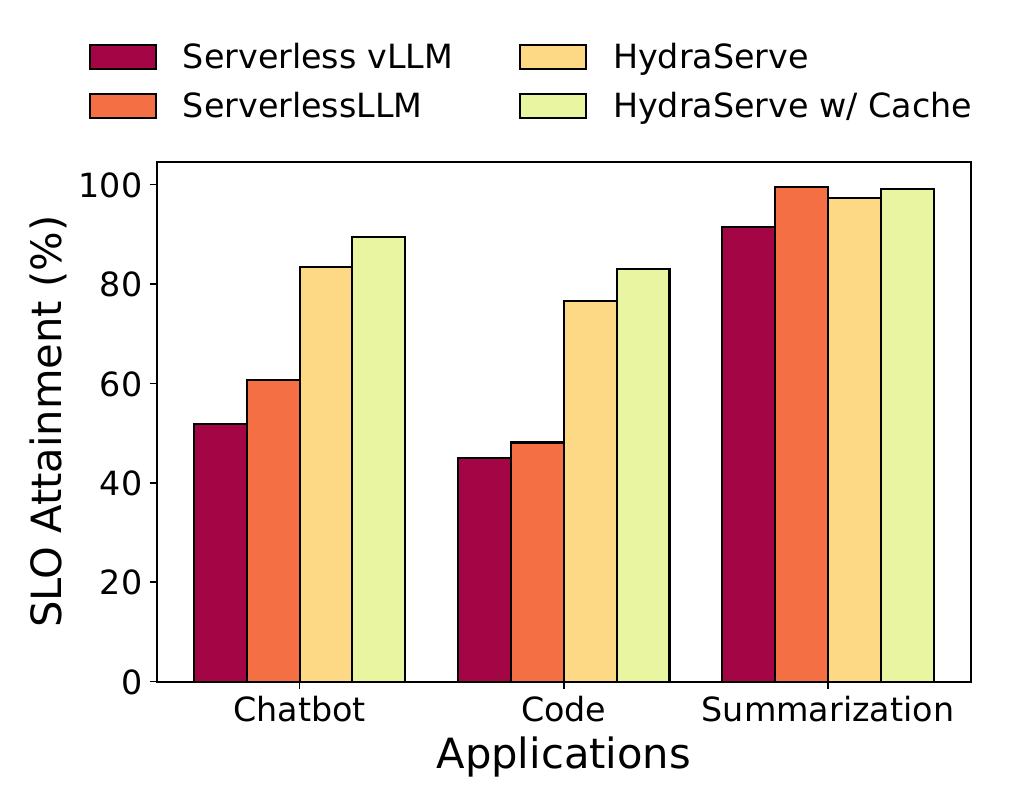}
        \vspace{-0.3in}
        \caption{TTFT SLO attainment for different applications.}
        \label{fig:eval-app}
    \end{minipage}
    \begin{minipage}[t]{0.49\linewidth}
        \centering
        \includegraphics[width=0.99\linewidth]{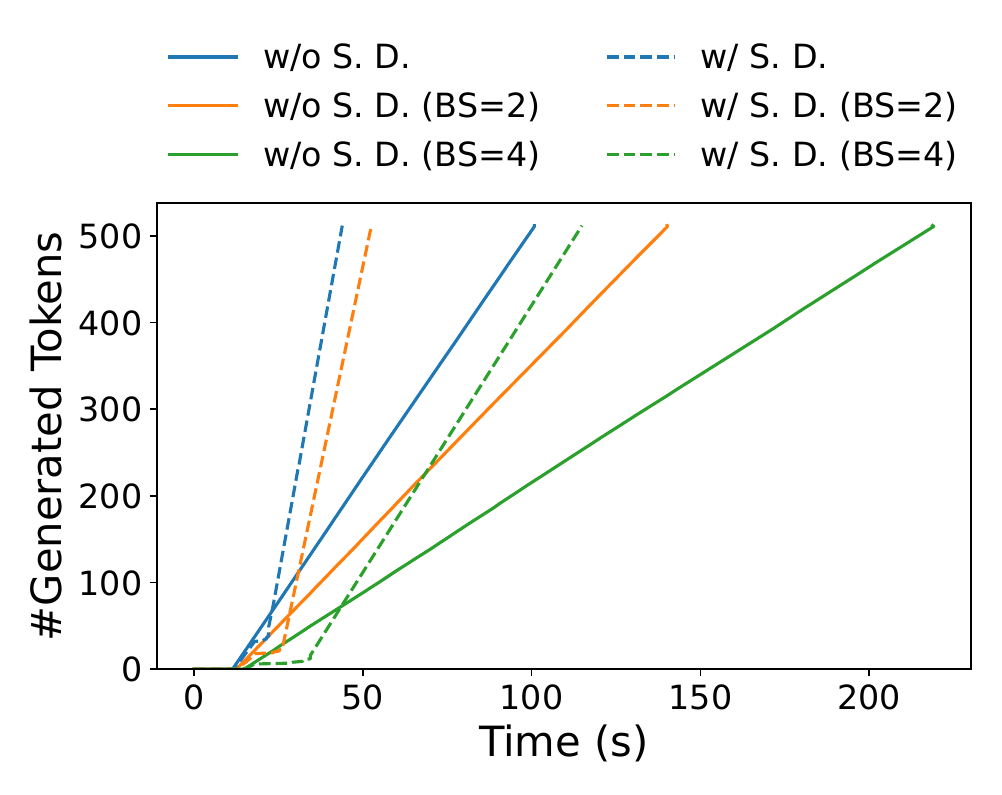}
        \vspace{-0.3in}
        \caption{Total tokens generated over time for different inputs.}
        \label{fig:eval-single}
    \end{minipage}
    \vspace{-1.5em}
\end{figure}

\begin{figure}[t]
    \centering
    \subfloat[TPOT ratios for different models.]
    {\includegraphics[width=0.47\linewidth]{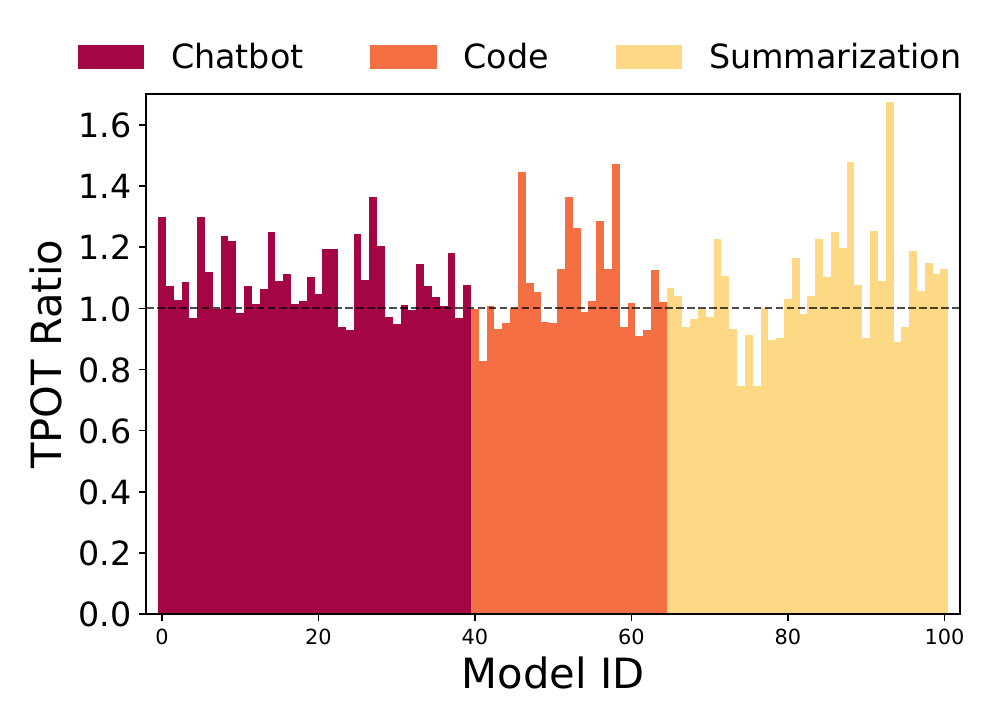}}
    \quad
    \subfloat[Cost ratios for different models.]
    {\includegraphics[width=0.47\linewidth]{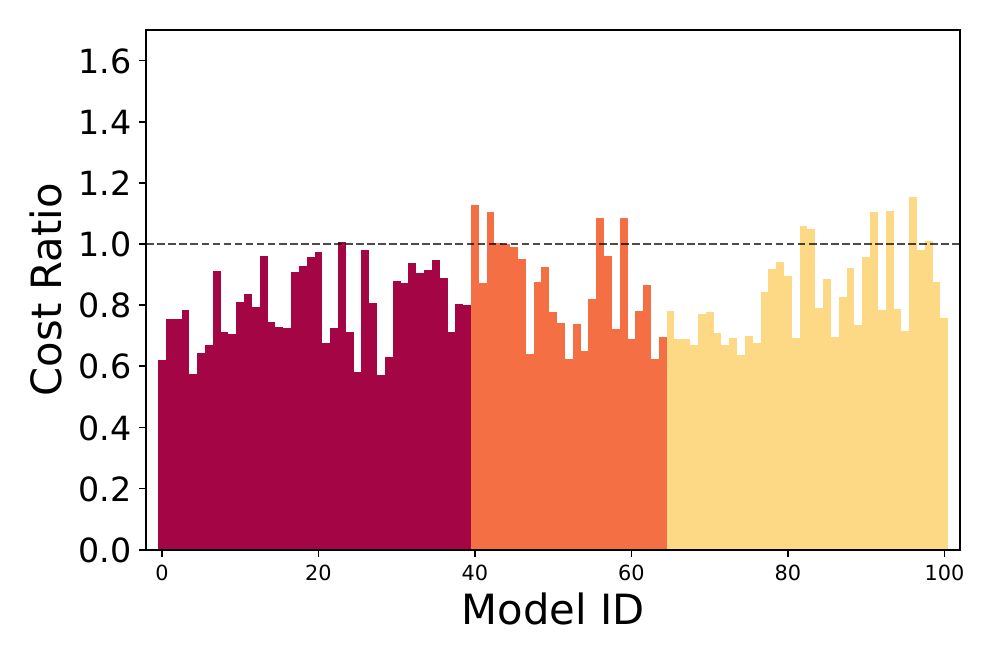}}
    \vspace{-0.1in}
    \caption{Relative TPOT and cost ratios (\sysname vs. serverless vLLM) for different models.}
    \label{fig:eval-detail}
    \vspace{-1em}
\end{figure}

\begin{figure}[t]
    \centering
    \vspace{-1em}
    \subfloat[Average TTFT of different loads.]
    {\includegraphics[width=0.47\linewidth]{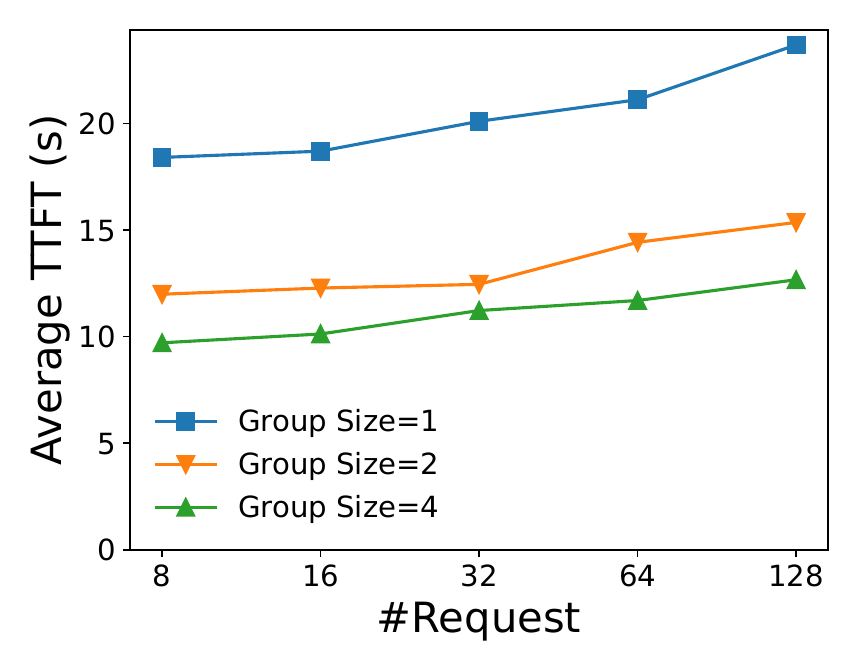}}
    \quad
    \subfloat[Average TPOT of different loads.]
    {\includegraphics[width=0.47\linewidth]{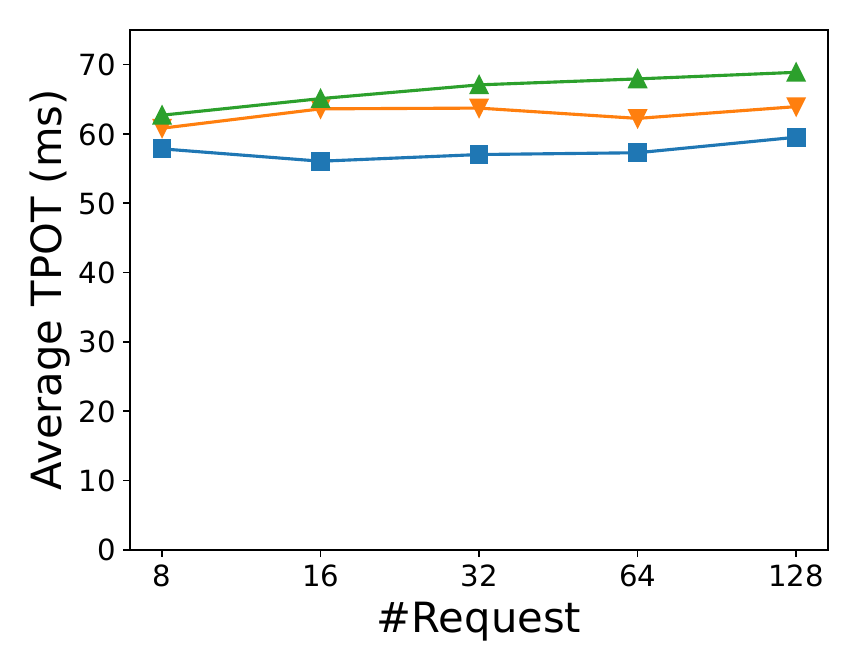}}
    \vspace{-0.1in}
    \caption{Performance comparison for handling bursty loads with different parallel group sizes.}
    \label{fig:eval-burst}
    \vspace{-1em}
\end{figure}

\parabf{Application analysis.}
Figure~\ref{fig:eval-app} illustrates the TTFT SLO attainment of chatbot, code completion, and summarization applications under CV=8 and RPS=0.6.
First, the results show that \sysname significantly enhances the TTFT SLO attainment for chatbot and code applications,
with up to 1.61$\times$ and 1.70$\times$ improvement, respectively.
The code application has lower SLO attainment compared to others because code completion tasks (i.e., requests in HumanEval dataset~\cite{humaneval}) have shorter average output length than chat tasks (i.e., requests in ShareGPT dataset~\cite{sharegpt})~\cite{distserve}.
Therefore, workers for code completion models keep alive for shorter time, leading to more cold starts.
Since the summarization application has loose SLOs, it has few SLO violations for all systems.

\parabf{TPOT and resource usage penalties.}
As discussed in \S\ref{sec:pipeline}, pipeline parallelism increases the worst-case TPOTs,
whereas allocating additional resources helps to reduce them.
We evaluate \sysname's TPOT and resource usage of different models to examine these trade-offs.
Figure~\ref{fig:eval-detail} illustrates the relative TPOT and cost ratios of \sysname compared to serverless vLLM under CV=8 and RPS=0.6.
The cost is proportional to the GPU memory-time product.
Regarding TPOT, \sysname exhibits only a 1.06$\times$ average increase compared to serverless vLLM,
with most increases occurring in chatbot and code models that have stringent TTFT requirements.
The TPOT penalty is modest because \sysname can quickly merge pipeline groups into single workers,
limiting performance degradation to only the first few tokens.
For some models, \sysname achieves shorter TPOT due to GPU performance fluctuations.

Figure~\ref{fig:eval-detail}(b) reveals that, surprisingly, \sysname consumes lower cost for most models.
This is because (1) for models with additional resource consumption, \sysname can quickly merge pipeline groups, and (2) \sysname enables faster worker startup, thereby reducing GPU usage during cold starts.
On average, \sysname reduces costs by 1.12$\times$ compared to serverless vLLM.

\subsection{Pipeline Consolidation}
\label{sec:eval-consolidation}

To evaluate the effectiveness of pipeline consolidation,
we measure \sysname's performance with two scaling methods.
We deploy Llama2-13B on V100 servers in testbed (i) and set the input and output length of each request to 512 tokens.

\parabf{Scaling down.}
We demonstrate the benefits of scaling down by presenting the generation time of each token.
We use different batch sizes and set pipeline parallelism size to 4.
As shown in Figure~\ref{fig:eval-single},
with scaling down, the system loads the remaining parts of the model in parallel with inference,
and migrates the key-value cache of the ongoing request once loading has completed.
This allows subsequent tokens to be generated at a faster speed.
As a result, scaling down reduces the end-to-end generation time by 1.90$\times$--2.67$\times$,
while maintaining almost same inference speeds during the early cold-start period.

\parabf{Scaling up.}
We evaluate scaling up by measuring \sysname's performance under bursty workloads.
We set the maximum batch size for each worker to 8 and vary the number of incoming requests.
Figure~\ref{fig:eval-burst}(a) illustrates the average TTFT across different loads.
The results show that larger pipeline parallelism sizes significantly reduce TTFT,
enabling the system to increase throughput earlier.
For example, when handling 128 concurrent requests (the maximum load for 16 V100 GPUs),
using four workers in a pipeline parallelism group reduces the average TTFT by 1.87$\times$.
Furthermore, Figure~\ref{fig:eval-burst}(b) indicates that scaling up incurs little inference performance overhead,
with the average TPOT increasing by 1.08$\times$--1.19$\times$.
This increase is attributed to the transmission overhead of intermediate results.

\begin{figure}[t]
    \centering
    \includegraphics[width=0.98\linewidth]{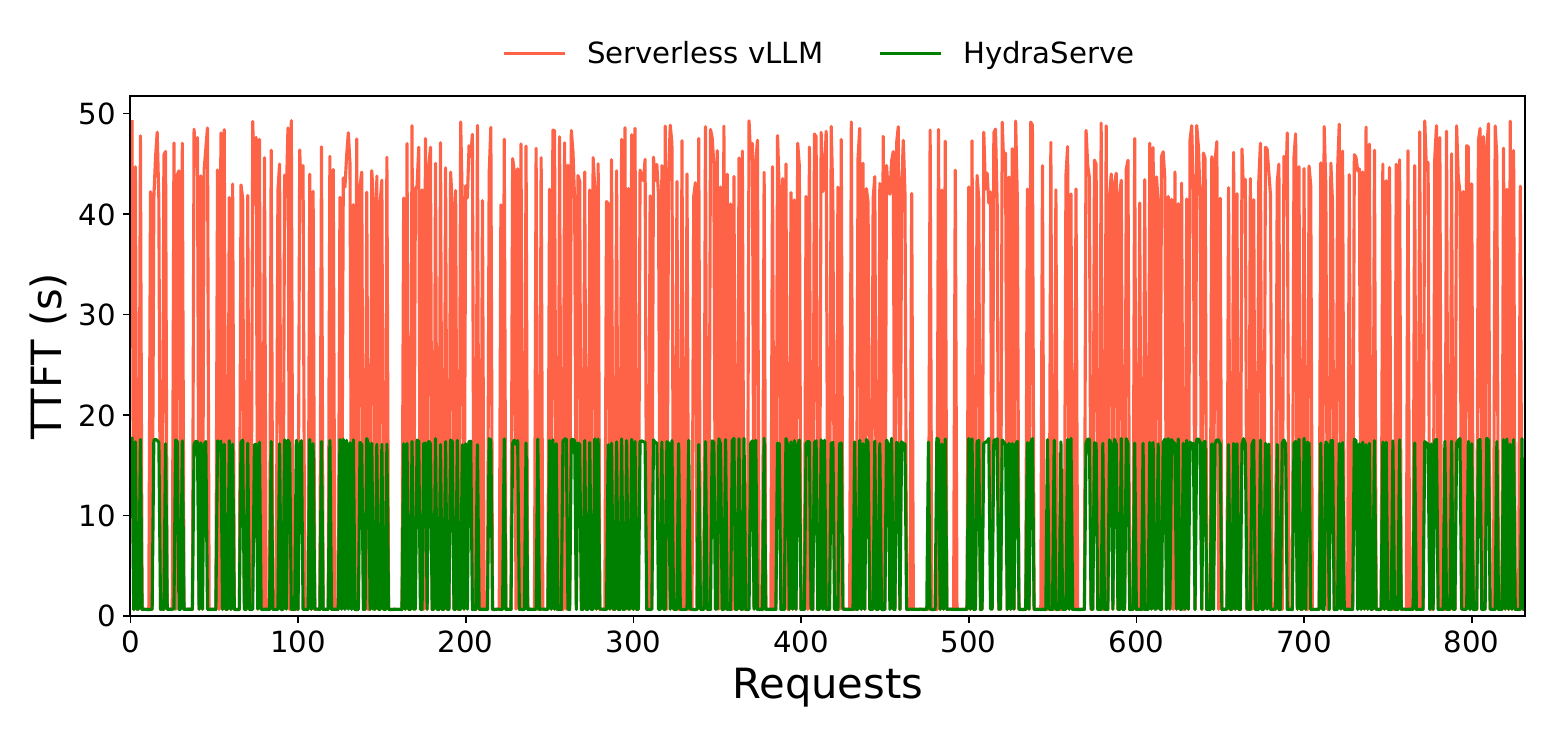}
    \vspace{-0.15in}
    \caption{TTFT of requests in brownfield evaluation.}
    \label{fig:eval-deployment}
    \vspace{-1em}
\end{figure}

\subsection{Brownfield Evaluation}
\label{sec:deployment}

We evaluate the prototype of \sysname in our production environment.
In this environment, functions cannot establish direct TCP connections with one another due to security constraints,
so we leverage a shared object in remote storage to enable inter-worker communication.
In evaluation, we use vLLM to run Llama2-7B on NVIDIA A10 GPUs with 24 GB GPU memory,
with requests generated following Microsoft Azure Function Trace~\cite{azure_trace}.
Figure~\ref{fig:eval-deployment} compares the cold start latency of \sysname to serverless vLLM.
The results demonstrate that \sysname significantly reduces cold start latency versus the original vLLM system,
achieving an average 2.6$\times$ reduction in cold-start TTFT.
This substantial improvement stems from \sysname's parallelized model fetching and optimized worker initialization processes.

 \section{Related Work}
\label{sec:related}

\parabf{Cold-start optimizations in serverless model serving.}
There have been plenty of works on reducing the cold start latency in serverless model serving~\cite{infaas, infless, asyfunc, faaswap, serverlessllm, medusa, blitzscale, deepflow}.
For instance, FaaSwap~\cite{faaswap} and ServerlessLLM~\cite{serverlessllm} cache models in local memory or SSDs,
while INFless~\cite{infless} and InstaInfer~\cite{socc24-prewarm} prewarm instances according to historical request patterns.
DeepFlow~\cite{deepflow} combines model caching and NPU forking to optimize cold start latencies.
BlitzScale~\cite{blitzscale} and $\lambda$Scale~\cite{lambdascale} achieve fast autoscaling using high-capacity networks,
while Medusa~\cite{medusa} speeds up CUDA graph and KV cache construction through state materialization.

\sysname addresses the cold-start problem in public clouds.
Its advantages are twofold.
First, it imposes no requirement on the hotness of models and effectively tackles the cold-start problem without relying on caching or worker cooperation.
Second, it reduces the end-to-end runtime preparation overhead without any offline profiling.

\parabf{LLM serving optimizations.}
Many LLM serving optimizations have been proposed to improve serving performance~\cite{li2023alpaserve, fastserve, loongserve, distserve, orca, vllm, infinigen, llumnix, sarathi}.
In the area of request scheduling, Orca~\cite{orca} introduces iteration-level scheduling to run more requests in parallel,
while Llumnix~\cite{llumnix} employs runtime scheduling to meet user SLOs.
For key-value cache management,
vLLM~\cite{vllm} leverages the concept of virtual memory in operating systems and InfiniGen~\cite{infinigen} optimizes KV block placement to improve inference efficiency.
Additionally, DistServe~\cite{distserve} disaggregates prefill and decoding phases to mitigate their performance interference.
\sysname specifically focuses on reducing cold start latency in serverless LLM serving and can easily integrate these optimizations.

\parabf{Pipeline parallelism.}
Pipeline parallelism has been widely used to scale GPU resources for training large models~\cite{huang2019gpipe, terapipe, narayanan2019pipedream, nips18_pipesgd, nips22_sapipe, megascale, graphpipe}.
Recently, researchers have also explored its usage during inference~\cite{li2023alpaserve, hpipe, pipeinfer, pipeedge,lambdascale, blitzscale}.
While AlpaServe~\cite{li2023alpaserve} observes that model parallelism improves GPU utilization under load spikes and develops a model placement policy,
HPipe~\cite{hpipe} and PipeEdge~\cite{pipeedge} apply pipeline parallelism to improve the inference performance for LLMs on edge devices.
$\lambda$Scale~\cite{lambdascale} also adopts pipeline parallelism to allow scaled-up workers to start inference earlier.
\sysname utilizes pipeline parallelism in a proactive manner that
creates multiple workers even if the model only needs one.
It further introduces pipeline consolidation to merge pipeline workers during inference. \section{Conclusion}
\label{sec:conclusion}
This paper presents \sysname,
a serverless LLM serving system designed to reduce cold start latency in public clouds.
\sysname targets the most time-consuming stages of a cold start: model fetching and runtime preparation.
To optimize model fetching, \sysname proactively distributes models across multiple servers,
alleviating the burden on any single server.
Furthermore, \sysname incorporates pipeline consolidation that merges workers back into individual endpoints to ensure efficient resource usage and high performance for warm requests.
For runtime preparation, \sysname accelerates the local worker startup by overlapping distinct stages.
Evaluation results show that \sysname reduces cold start latency by 1.7$\times$-- 4.7$\times$ compared to baselines and improves TTFT SLO attainment by 1.43$\times$--1.74$\times$ under various constraints.
By integrating these optimizations, \sysname offers a robust and efficient solution to meet user-defined SLOs in serverless LLM serving, especially for long-tail models.
 
\parabf{Acknowledgments.} We sincerely thank our shepherd Arvind Krishnamurthy
and the anonymous reviewers for their valuable feedback on this paper. This work
was supported in part by the National Key Research and Development Program of
China under Grant 2022YFB4500700, the Scientific Research Innovation Capability
Support Project for Young Faculty under Grant ZYGXQNJSKYCXNLZCXM-I1, the
Fundamental Research Funds for the Central Universities, Peking University, and
the National Natural Science Foundation of China under Grant 62172008 and
62325201. Xin Jin is the corresponding author. Chiheng Lou, Sheng Qi, Chao Jin,
Xuanzhe Liu, and Xin Jin are also with the Key Laboratory of High Confidence
Software Technologies (Peking University), Ministry of Education. \label{lastpage}

{\bibliographystyle{ieeetr}
\bibliography{xin}}

\clearpage
\appendix
\onecolumn
\section{TPOT SLO Attainment}

\begin{figure*}[htbp]
    \centering
    \includegraphics[width=0.99\textwidth]{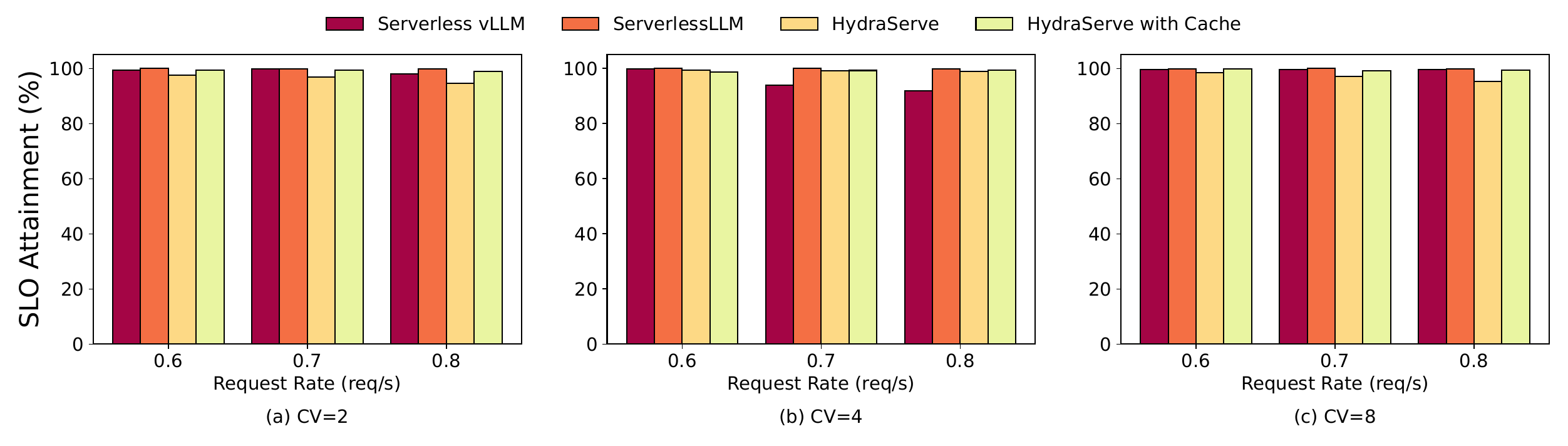}
    \vspace{-0.2in}
    \caption{TPOT SLO attainment of systems under different CVs.}
    \label{fig:eval-end2end-tpot}
    \vspace{-1em}
\end{figure*}

We provide the TPOT SLO attainment of systems under different CVs in Figure~\ref{fig:eval-end2end-tpot}.
All systems achieve over 95\% TPOT SLO attainment in most scenarios,
and more than 90\% under all CVs and RPS configurations.

\end{document}